\documentclass[a4paper,12pt]{article}

\usepackage[utf8]{inputenc} 
\usepackage{color} 
\usepackage{graphicx}       
\usepackage{subcaption}

\usepackage{amsmath}  
\usepackage{amssymb}  
\usepackage{bm}       
\usepackage{mathtools}

\usepackage{cite}  
\usepackage[linktoc=all]{hyperref}
\usepackage{tocbibind} 

\usepackage{float}

\graphicspath{{images/}} 

\newcommand{\ve}{\vec}
\newcommand{\br}[1]{\left(#1\right)}
\newcommand{\sbr}[1]{\left[#1\right]}
\newcommand{\grad}{\nabla}
\newcommand{\ti}[1]{\tilde{#1}}

\renewcommand{\a}{\alpha}

\renewcommand{\d}{\delta}
\newcommand{\f}{\phi}
\newcommand{\m}{\mu}
\newcommand{\n}{\nu}
\renewcommand{\r}{\rho}
\newcommand{\s}{\sigma}

\newcommand{\p}{\pi}
\renewcommand{\k}{\kappa}
\newcommand{\g}{\gamma}
\renewcommand{\t}{\tau}
\newcommand{\w}{\omega}

\newcommand{\D}{\Delta}

\renewcommand{\P}{\Pi}

\numberwithin{equation}{section}

\allowdisplaybreaks

\begin{document}

\title{ {\bf Gravitational Collapse in Cubic Horndeski Theories}}
\author{Pau Figueras and Tiago França }
\date{}
\maketitle

\thispagestyle{empty}

\begin{center}
\textit{ School of Mathematical Sciences, Queen Mary University of London}\\ \textit{ Mile End Road, London, E1 4NS, United Kingdom}
\vspace{0.5cm}
\\
\texttt{p.figueras@qmul.ac.uk}, \texttt{t.e.franca@qmul.ac.uk}
\end{center}

\vspace{1cm}
\begin{abstract}   
We study spherically symmetric gravitational collapse in cubic Horndeski theories of gravity. By varying the coupling constants  and the initial amplitude of the scalar field, we determine the region in the space of couplings and amplitudes for which it is possible to construct global solutions to the Horndeski theories. Furthermore, we identify the regime of validity of effective field theory as the sub-region for which a certain weak field condition remains small at all times. We evolve the initial data using the CCZ4 formulation of the Einstein equations and horizon penetrating coordinates without assuming spherical symmetry.  

\end{abstract}

\newpage

\thispagestyle{empty}
\setcounter{footnote}{0}
\setcounter{page}{0}

\tableofcontents
\newpage

\section{Introduction and Summary}\label{intro}

The detections of gravitational waves produced in mergers of compact objects  \cite{Abbott:2016blz,TheLIGOScientific:2017qsa} have revolutionised the field of gravitational physics, giving rise to the era of gravitational wave astronomy. Thanks to recent upgrades of the detectors, gravitational waves detections are made almost on a weekly basis. Therefore, we now have an unprecedented amount of data that gives us access to the strong field regime of gravity. The situation is only going to get better in the future, with new detectors gradually added to the network in the coming years and a forthcoming third generation of  detectors such as the Einstein Telescope and ultimately Lisa, a space-based observatory. Therefore, very soon we will enter the era of precision gravitational wave astronomy.  

These advancements offer the opportunity (and carry the duty) to test Einstein’s theory of general relativity (GR) using gravitational waves. One of the main challenges in doing these tests is to come up with templates of waveforms in alternative theories of gravity. One possibility is to focus on those phases of the binary that can be treated using perturbation theory, namely the inspiral \cite{Berti:2018cxi} and the ringdown phases \cite{Berti:2018vdi,Cano:2020cao} respectively. However, the present data suggests that the corrections to GR are small. Therefore, one may hope that there is a better chance to detect some deviations from GR in the strong field regime, namely in the merger phase, where some effects may be enhanced. This would be the case for deviations from GR that are sourced by spacetime curvature, such as higher derivative corrections. So far the merger phase has been modelled phenomenologically \cite{Yunes:2009ke,Agathos:2013upa}, or by treating the deviations from GR perturbatively \cite{53_dcSgravity_bounds_on_parameter,109_EdGB_binaries,52_dCSgravity_BBH_non_degenerate_with_GR,111_EdGB_binaries_2,110_dCS_binaries}; only the so called scalar-tensor and scalar-vector-tensor theories of gravity have been considered in their full non-linear glory in all phases of the binary \cite{Healy:2011ef,Barausse:2012da,51_EMDtheoryNOTgood,Sagunski:2017nzb}.

Another difficulty is that there are many alternative theories of gravity, and each one of them modifies GR in a different way: adding new fields, breaking some symmetries, adding new terms to the action, etc.. At the moment there is no theoretical consensus nor any experimental evidence that favours a particular theory. Each modification of GR should be reflected in a unique way in the corresponding waveforms and hence the interest in analysing gravitational waves in alternative theories of gravity.  However, in many of these theories it is not known whether the initial value problem is well-posed. Without a well-posed initial value problem, one cannot possibly simulate the non-linear regime of the theory on a computer and obtain the desired waveforms.  There have been some recent efforts that have successfully managed to construct well-posed formulations of certain modified theories of gravity of physical interest \cite{27_aron,85_AronMofifiedHarmonicGauge,97_Aron_ModifiedHarmonicGauge_FULL}.\footnote{Earlier works studided the well-posedness of Lovelock and Horndeski theories and found that the equations of motion are weakly hyperbolic in a certain class of generalised harmonic gauges \cite{42_Harvey_LovelockAndHorndeski_Hyperbolicity,19_horndeski_papallo}.} Alternatively, \cite{Cayuso:2017iqc,Allwright:2018rut} have proposed to find well-posed formulations of alternative theories of gravity extending the M\"uller-Israel-Stewart formalism of viscous relativistic hydrodynamics \cite{Muller:1967aa,Israel:1976tn,Israel:1976213,Israel:1979wp} to those theories of gravity. Very recently \cite{Cayuso:2020lca} succeeded in applying this formalism to theories of gravity with higher curvature corrections assuming spherical symmetry.

Treating the modifications to Einstein’s gravity perturbatively may seem justified given that the present data indicates that they are small. In this case, there are no issues with the well-posedness of the equations and this is the approach that has been adopted in a number of papers \cite{53_dcSgravity_bounds_on_parameter,109_EdGB_binaries,52_dCSgravity_BBH_non_degenerate_with_GR,111_EdGB_binaries_2,Cano:2020cao,deRham:2020ejn}. However, it has some serious limitations: it is well-known that small effects can accumulate over time and eventually lead to a breakdown of perturbation theory in a regime where it should still be valid. Furthermore, this approach is completely insensitive to certain non-perturbative effects encoded in the full non-linear theory. For instance, the non-linear perturbation theory around anti-de Sitter space breaks down precisely before a black hole forms \cite{Bizon:2011gg}. 

GR is a classical theory and, as such, it should be understood as low energy effective field theory (EFT) of gravity. Indeed, on general grounds, one expects that at sufficiently small distances, Einstein’s theory will be modified by quantum corrections. From the point of view of EFT, these corrections can be organised in a series expansion involving increasing powers of the curvature tensor, and consequently higher derivatives of the spacetime metric.  Since in current experiments we are only probing gravity at low energies, we should only be sensitive to a finite number of terms in the otherwise infinite series of corrections to GR. Moreover, the details of the UV completion of gravity should not be important at such low energies. Higher derivative corrections are just one example of the myriad of possible modifications to GR that have been considered.  Any of these alternative theories of gravity should be understood as truncated low energy EFT and, as such, they only make sense if the corrections to GR are small.
   
One particular modified theory of gravity which is known to have a well-posed initial value problem is Horndeski theory \cite{27_aron,85_AronMofifiedHarmonicGauge,97_Aron_ModifiedHarmonicGauge_FULL}.\footnote{Reference  \cite{Rendall:2005fv} had previously proven well-posedness of the initial value problem for the so called $k$-essence theories, which are a subclass of the Horndeski theories considered in these papers.} This is the most general theory of a metric tensor coupled to a scalar field  with second order equations of motion arising from a diffeomorphism invariant action in four spacetime dimensions.\footnote{This theory was  first found by Horndeski \cite{69_Horndeski_Original_1974} and rediscovered in other works \cite{122_re_discovery_of_Horndeski_Galileon,123_Post_analysis_of_re-discovery_of_Horndeski_Galileon,124_Post_analysis_of_re-discovery_of_Horndeski_Galileon_more}.} The general action for this theory is\footnote{The Teleparallel gravity version of this theory has been recently worked out in \cite{Bahamonde:2019shr,Bahamonde:2019ipm}. While this version may offer a phenomenologically attractive avenue to explore, the well-posedness of the initial value problem in these theories has not been established.}
\begin{equation}\label{eq:Horndeski_general}
    \mathcal{S} := \frac{1}{\k}\int dx^4 \sqrt{-g} \br{\mathcal{L}_1+\mathcal{L}_2+\mathcal{L}_3+\mathcal{L}_4+\mathcal{L}_5},
\end{equation}
with
\begin{equation}
    \begin{aligned}
        \mathcal{L}_1 =&~ R + X - V(\f)\,,\\
        \mathcal{L}_2 =&~ G_2(\f,X)\,,\\
        \mathcal{L}_3 =&~ G_3(\f,X)\,\square\f\,,\\
        \mathcal{L}_4 =&~ G_4(\f,X)\,R + \partial_X G_4(\f,X)\,\sbr{\br{\square\f}^2-\br{\grad_\m\grad_\n\f}\br{\grad^\m\grad^\n\f}}\,,\\
        \mathcal{L}_5 =&~ G_5(\f,X)G_{\m\n}\grad^\m\grad^\n\f  \\
        & - \frac{1}{6}\partial_X G_5(\f,X)\,\sbr{\br{\square\f}^3 -3\square\f\br{\grad_\m\grad_\n\f}\br{\grad^\m\grad^\n\f} + 2\br{\grad_\m\grad_\n\f}\br{\grad^\n\grad^\r\f}\br{\grad_\r\grad^\m\f}}\,,
    \end{aligned}
\end{equation}
where $\k:=16\p G$ is related to the 4-dimensional  Newton's constant; $\f$ is a scalar field and $X := -\tfrac{1}{2}(\grad_\mu\phi)(\grad^\mu\phi)$; $G_i$ ($i=2,3,4,5$) are freely specifiable functions, and $R$ and $G_{\m\n}$ are the Ricci scalar and Einstein tensor of the spacetime metric $g_{\m\n}$, respectively. Having only second-order equations is essential to avoid Ostrogradsky instabilities \cite{83_Ostrogradsky,84_Ostrogradsky2}. This theory has found numerous applications to cosmology; the literature on the subject is vast and we will not attempt to review it here. We refer the reader to the recent reviews \cite{44_ModifiedGravityReview,Kobayashi:2019hrl,93_ModifiedGravity_Summary}.
In this work we study the non-linear regime of a subclass of Horndeski theories for which \cite{27_aron} found a well-posed CCZ4 formulation of the Einstein equations. In this paper, unlike \cite{53_dcSgravity_bounds_on_parameter,109_EdGB_binaries,52_dCSgravity_BBH_non_degenerate_with_GR,111_EdGB_binaries_2}, we consider the theory in its full non-linear baroque splendour, which allows us to explore its distinctive non-perturbative physics; our goal is to identify the weakly coupled regime of the theory so that it can be consistently treated as a valid EFT from which one can obtain meaningful predictions.  Rather than studying a specific phenomenologically viable theory, our ultimate goal is to identify general features in the waveforms that do not depend on the details and that can be attributed to the higher derivatives and non-linearities in the action. Therefore, we treat it as a toy model that can give us a glimpse of the  type of effects that one can expect in more complicated theories which involve higher derivatives of the spacetime metric tensor.  

For clarity of the presentation, we have split our work in a series of two articles, of which this is the first one. In this paper we study gravitational collapse and black hole formation in Horndeski theory. Our goal is to identify the region in the space of couplings for which the Horndeski theories under consideration are weakly coupled throughout the evolution. Using these results, in a companion paper we study black hole binary mergers, treating the theory fully non-linearly while remaining the regime of validity of EFT in all phases of the binary. In the following subsection, we summarise the main results in the present article, and refer the reader to  the companion paper \cite{PFTF2} for the results on black hole binaries. 

\subsection{Summary of the main results}\label{summary}

In this paper we consider gravitational collapse in Horndeski theories using as initial data a spherically symmetric lump of scalar field \eqref{eq:phiData}. Even though the initial data is spherically symmetric, we evolve it using a 3+1 evolution code based on \texttt{GRChombo} \cite{grchombo}, without symmetry assumptions. We have also considered gravitational collapse of some non-spherical scalar field configurations but we did not observe significant differences from the spherically symmetric case. However, a thorough study of gravitational collapse beyond spherical symmetry in Horndeski theories is beyond the scope of this paper.

Before we describe our results, we comment on previous works that are directly related to ours. Gravitational collapse and black hole dynamics in spherical symmetry in Einstein-dilaton-Gauss-Bonnet (EdGB) theory has been studied before \cite{25_pretorius_horndeski_19,39_Pretorius_EdGB_elliptic,73_pretorius_EdGB_elliptic_inside_BHs,Ripley:2020vpk}. This theory can be considered to be a member of the Horndeski class, but the mapping between the two is highly non-trivial \cite{98_GeneralizedGalileon_Horndeski_and_GB}. In these papers the authors study, among other things, the hyperbolicity of the equations of motion in various regions of the spacetime, including the interior of black holes, as a function of the coupling. They show that for large enough couplings the equations of motion can change character from hyperbolic to elliptic, even outside black holes, in which case one cannot solve them as an evolution problem. In a related work, \cite{38_horndeski_new_may_example_luis} considers the conditions under which one may be able construct global solutions of Horndeski theories. In this paper, the authors study in detail the hyperbolicity of the equations of motion and the pathologies that may arise during the evolution in some specific examples. They also perform  numerical simulations of spherically symmetric scalar field collapse to illustrate the breakdown of the hyperbolicity at strong coupling in different situations. Our work can be considered as an extension of these papers in different directions, as we now explain.

In this article we consider the so called cubic Horndeski theories \eqref{eq:action}, for which \cite{27_aron} showed that they have a well-posed initial value problem in the CCZ4 formulation of the Einstein equations and in puncture gauge. Because we are not particularly interested in a specific theory but rather in 
identifying general features of the non-linear dynamics of Horndeski theories, we consider two particularly simple and illustrative cases, see equation \eqref{eq:relevantGs}. In fact, from the point of view of EFT, the $G_2$ theory considered here, eq. \eqref{eq:relevantGs}, is the most general scalar matter term up to four derivatives that one can include to the action \cite{Weinberg:2008hq}. In order for these theories to make sense as EFTs, the Horndeski terms have to be suitably small compared to the GR terms. Indeed, the well-posedness result of \cite{27_aron} only holds if a certain weak field condition is satisfied. For the class of theories that we consider, the relevant weak field conditions are given by \eqref{eq:WFC_2}. The main goal of this paper  is to identify the region in the space of initial conditions and couplings for which the weak field conditions \eqref{eq:WFC_2} are small at all times.  

In our simulations of scalar field collapse we keep the radius $r_0$ and width $\omega$ of the initial Gaussian lump fixed, and vary both the amplitude $A$ and Horndeski coupling ($g_2$ or $g_3$ depending on the theory under consideration). For every pair $(A,g_2)$ or $(A,g_3)$, we monitor both the character of the equations of motion of the scalar field\footnote{The evolution equations for the metric are given by the CCZ4 equations which are (strongly) hyperbolic.} and the weak field conditions \eqref{eq:WFC_2} everywhere in spacetime, except in a certain region of the interior of black holes when they form.  It seems reasonable to accept the breakdown of EFT in a region sufficiently close to a singularity as long as this region is covered by a horizon. In this case, there is no loss of predictivity since this region is causally disconnected from the Universe outside the black hole, where EFT remains valid. The same criterion was adopted in \cite{73_pretorius_EdGB_elliptic_inside_BHs}.

\begin{figure}[h!]
\centering
\hspace*{-10mm}\includegraphics[width=1.15\textwidth]{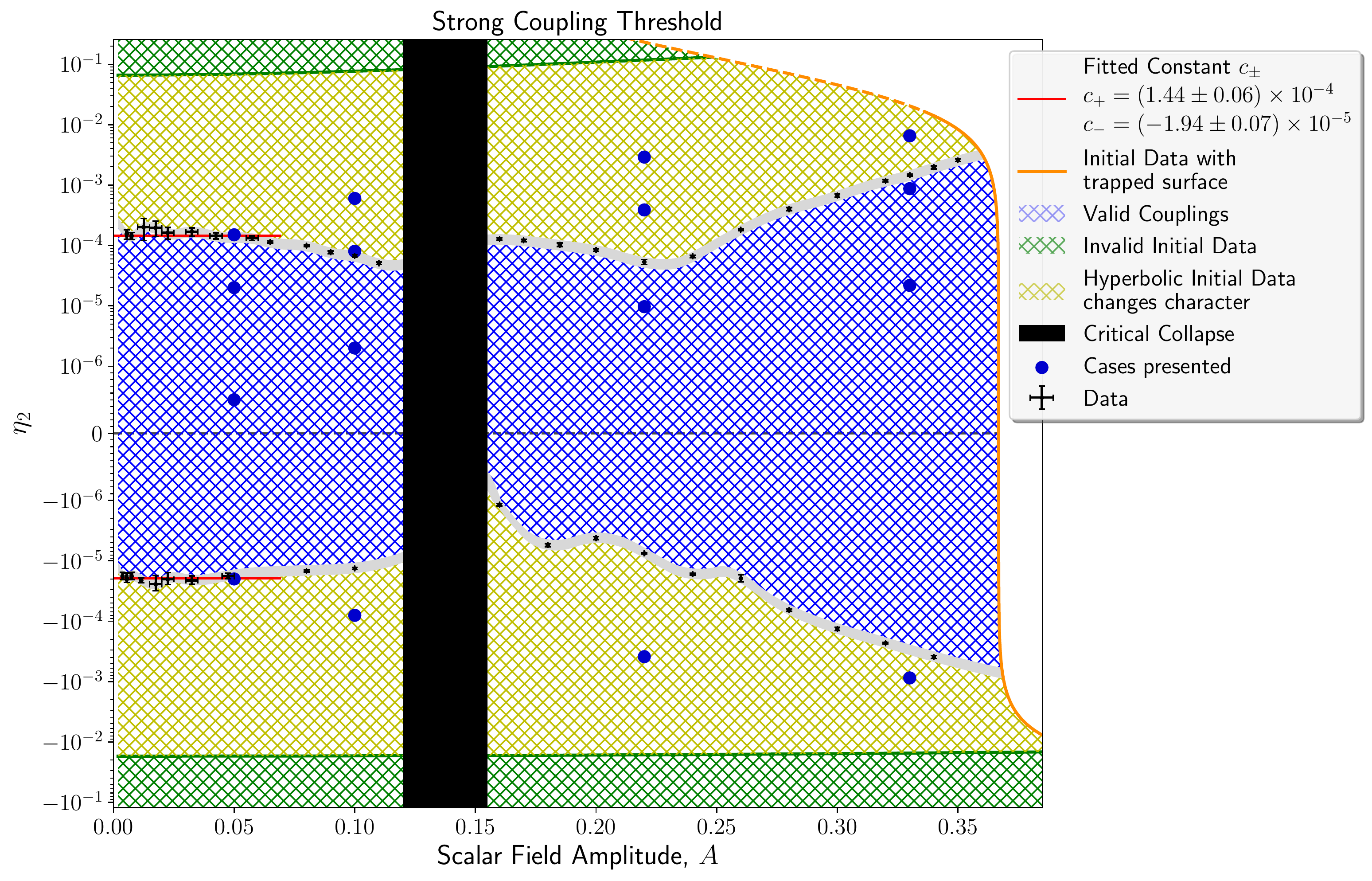}
\caption{Dynamical regimes of the $G_2 = g_2\,X^2$ theory as a function of the initial amplitude $A$ and the dimensionless coupling constant $\eta_2$, see eq. \eqref{eq:eta2}. The black band denotes the region near critical collapse; black holes form to the right of this band. The orange curve on the right marks the region where the initial data contains a trapped surface. The scalar equation is hyperbolic at all times in the blue region; EFT is valid in the interior of this region. In the yellow region, the scalar equation is initially hyperbolic but it changes character during the evolution. In the green region the initial value problem is not well-posed.}
\label{fig:results_eta2}
\end{figure}

Our main results for the $G_2\neq 0$, $G_3=0$ theory are summarised in Fig. \ref{fig:results_eta2}.  The analogous figure for the $G_3\neq 0$, $G_2=0$ theory is qualitatively similar and can be found in Section \ref{sec:G3_neq_0}, Fig. \ref{fig:results_eta3}. For the sake of definiteness, in the following we shall focus our discussion on the $G_2\neq 0$, $G_3=0$ theory but essentially the same conclusions apply to the $G_3\neq 0$, $G_2=0$ theory. 

The dimensionless coupling constants $\eta_2$ and $\eta_3$, see eqs. \eqref{eq:eta2}-\eqref{eq:eta3}, control the future development for our initial data; in other cubic Horndeski theories one should be able to define analogous dimensionless couplings, and therefore the conclusions of this paper should apply to those theories as well. In Fig. \ref{fig:results_eta2} we show the various dynamical regimes of the $G_2$ theory  as a function of  $\eta_2$ and the initial scalar amplitude $A$. As one would expect, the weakly coupled regime of the theory corresponds to  suitably small values of $\eta_2$, but the boundary of this region depends non-trivially on the scalar amplitude.

The blue region in Fig. \ref{fig:results_eta2} denotes the values of $(A,\eta_2)$ for which the scalar equation is hyperbolic at all times. The yellow region corresponds to the values of $(A,\eta_2)$ for which the scalar equation is initially hyperbolic, and hence the initial value problem is well-posed, but it changes character during the evolution, signalling a breakdown of the theory. The green region  corresponds to the values of $(A,\eta_2)$ for which the scalar equation is not hyperbolic on the initial data slice and hence the initial value problem is not well-posed. The black band in Fig. \ref{fig:results_eta2} corresponds to the range of amplitudes for which the future development of the initial data gets close to Choptuik's critical solution \cite{Choptuik:1992jv}, which is a naked singularity. This band splits the figure into two regions corresponding to the small and large data regimes: for initial data in  the blue region to the left of the black band, the scalar field disperses to infinity. On the other hand, initial data in the blue region to the right of the black band collapses into a black hole. 

For initial data in any of  blue regions in Fig. \ref{fig:results_eta2} it is possible to construct global solutions to the $G_2=g_2\,X^2$ Horndeski theory. Away from the boundary of this region, the deviations from GR are ``small" everywhere on and outside black holes (if there are any) for all times. By ``small" here we mean that the weak field condition \eqref{eq:WFC_2} is satisfied. Therefore, we identify the interior of the blue region as the regime of validity of EFT for the corresponding Horndeski theory. Of course, when black holes form during the evolution, EFT will break down near the singularity, just as GR does. In this case, we excise a portion of the interior of the black hole since it is causally disconnected from external observers.  For values of $(A,\eta_2)$ close to the boundary of the blue region, the weak field condition \eqref{eq:WFC_2}  can become $\mathcal{O}(1)$ during the evolution while the scalar equation remains hyperbolic. In this case one may argue that even though the theory has a well-posed initial value problem, higher derivative corrections not included in the action \eqref{eq:action} should become important and hence one should not trust the theory as it stands. 

As mentioned above, the yellow region in Fig. \ref{fig:results_eta2} denotes the values of $(A,\eta_2)$ for which the evolution breaks downs due to the change of character of the scalar equation and this breakdown cannot be hidden behind a horizon. The change of character of the scalar equation is typically associated  to the weak field conditions becoming $\mathcal{O}(1)$ or larger but this is not always the case. Indeed, for certain values of $(A,\eta_2)$, and in particular for $\eta_2<0$, the weak field condition can be $\mathcal{O}(10^{-2})$ during the evolution and yet the equations change character. Beyond this point it is no longer possible to solve the theory as an initial value problem. However, we note that whenever the equations change character, the weak field condition is much larger than the dimensionless coupling $\eta_2$ determined from the initial data. Therefore, in a certain sense, the theory becomes strongly coupled right before it breaks down.  In Section \ref{numerics}  we study in detail how and where in spacetime the loss of hyperbolicity of the scalar equation happens depending on the Horndeski couplings and we correlate it to the weak field conditions \eqref{eq:WFC_2}.  For $0<A\lesssim 0.05$, the boundary between the blue and yellow regions  is given by a constant value of $\eta_2\sim (1.44\pm0.06)\times 10^{-4}$ and  $\eta_2\sim(-1.94\pm 0.07)\times 10^{-5}$ respectively. This is non-trivial since the location of this boundary is obtained from the non-linear evolution of the initial data.  As we will see in Section \ref{sec:G2_neq_0}, for $\eta_2>0$ the breakdown of the evolution happens through a Tricomi-type-of transition while for $\eta_2<0$ the transition is of the Keldysh type.

A zoom in of Fig. \ref{fig:results_eta2} near the black band would show that $\eta_2\to 0$ as one approaches the critical regime from both sides. This is expected since for $A$ near the critical amplitude $A_\ast=0.13\pm 0.01$, the gradients of both the metric tensor and the scalar field become very large as the solution approaches the critical solution, which leads to a change of character of the scalar equation unless $g_2\to 0$ as $A\to A_\ast$.  Since the regime of validity of EFT is essentially the empty set at the critical solution, in the rest of the paper we will purposely avoid the region near criticality.\footnote{Reference \cite{Gannouji:2020kas} studies critical collapse in $k$-essence models. We thank Eugeny Babichev for bringing our attention to this article.} For values of $A>A_\ast$, a black hole forms during the evolution of the initial data. The larger the value of $A$, the larger the black hole that forms and the sooner it forms. Since larger black holes result in lower curvatures on the horizon scale,  larger values of the couplings are allowed and yet the theory remains weakly coupled on and outside the black hole. This is the reason why $\eta_2$ increases for larger $A$.  For sufficiently large $A$, the initial data already contains a trapped surface. Since we are interested in studying gravitational collapse, we do not consider those values of $A$.

It is clear from the previous discussion that our weak field conditions \eqref{eq:WFC_2} bear some relation with the hyperbolicity condition of the scalar equation of motion \eqref{eq:scalar} but such a relation is not a direct one. It is possible that one can come up with  refined and sharp weak field conditions that also capture the change of character of the equations when they are violated but finding them is beyond the scope of this paper. It follows from our analysis that the regime of validity of EFT corresponds to the weak field conditions \eqref{eq:WFC_2} being satisfied (to justify that higher derivative terms in \eqref{eq:action} can be neglected) \textit{and} that the initial value problem is well-posed, i.e., the scalar equations of motion are hyperbolic everywhere in spacetime, perhaps except in a small region inside black holes. These two conditions are satisfied in the interior of the blue region in Figs. \ref{fig:results_eta2} and \ref{fig:results_eta3}. For initial data in this region, the Horndeski theories that we have considered are valid EFTs and global solutions can be constructed. We note that whilst the conditions for hyperbolicity and the weak field conditions \eqref{eq:WFC_2} overlap near the GR limit, the latter are not necessarily contained in the former far away from GR.\footnote{We thank Harvey Reall for discussions on this issue.}

The rest of the paper is organised as follows. In Section \ref{cubic} we present the theories that we consider and we analyse the corresponding hyperbolicity conditions. In Section \ref{numerics} we present and analyse the results of our numerical simulations. Subsection \ref{sec:G2_neq_0} discusses in detail the dynamics of the $G_2\neq 0$ theories, while the $G_3\neq 0$ theories are dealt with in Subsection \ref{sec:G3_neq_0}. We conclude with some final remarks in Section \ref{conclusions}. We have relegated some technical details to the Appendices. In Appendix \ref{appendix:ccz4} we write down the equations of motion for scalar field  and the effective scalar metric in a 3+1 form.  We collect some technical results in Appendices \ref{appendix:det_eff_metric} and \ref{appendix:excision}, and  the convergence tests are presented in Appendix \ref{appendix:convergence}. Appendix \ref{appendix:other} contains the results of certain numerical simulations that are also relevant for the main text. In this paper we adopt the following notation; we use Greek letters ($\m$, $\n$, $\r$, ...) to denote full spacetime indices and Latin letters ($i$, $j$, $k$, ...) for the spatial ones. We adopt the mostly plus metric signature, and we set $G=c=1$.

\section{Cubic Horndeski Theories}\label{cubic}
%

\subsection{Equations of motion}
\label{sec:eoms}
%
In this paper we consider the special subset of Horndeski theories for which \cite{27_aron} proved well-posedness of the initial value problem in both the BSSN and CCZ4 formulations of the Einstein equations in the usual gauges used in numerical relativity. This class of theories is given by setting $G_4=G_5=0$ in the general Horndeski action \eqref{eq:Horndeski_general}. 
This results in the so called cubic Horndeski theories described by the action
\begin{equation}\label{eq:action}
    \mathcal{S} := \frac{1}{\k}\int dx^4 \sqrt{-g} \big[R + X - V(\f) + G_2(\f , X) + G_3(\f , X)\square \f\big].
\end{equation}
Here, $X=-\frac{1}{2}(\nabla_\m\f)(\nabla^\m\f)$ and $V(\f)$ are the usual kinetic and potential terms respectively in the standard action for a minimally coupled scalar field, and $G_2(\f,X)$ and $G_3(\phi,X)$ are arbitrary functions of their arguments. In this paper, we have explicitly separated the canonical kinetic and potential terms from $G_2$ so that $G_2$ and $G_3$ parametrise the higher derivative terms and non-minimal couplings of the scalar field to gravity. The resulting Einstein equations are:
\begin{align}\label{eq:einstein}
        G_{\m\n} =&~ g_{\mu\nu} \bigl(G_2 + X - V + 2\, X\, \partial_\phi G_3\bigr) +(\nabla_{\mu}\phi)(\nabla_{\nu}\phi) \bigl(1 + \partial_X G_2 + 2\, \partial_\phi G_3\bigr)\\
        &+\partial_X G_3\bigl[(\square\phi)(\nabla_{\mu}\phi) (\nabla_{\nu}\phi) - 2\,(\nabla^{\rho}\phi)(\nabla_{(\mu}\phi) \nabla_{\nu)}\nabla_{\rho}\phi  + g_{\mu\nu} (\nabla^{\rho}\phi)(\nabla^{\sigma}\phi) \nabla_{\rho}\nabla_{\sigma}\phi \bigr],\nonumber
\end{align}
where $G_{\m\nu}$ is the Einstein tensor. The equation of motion for the scalar field is:\footnote{The direct variation of the action with respect to the scalar field yields a term $\partial_X G_3\, R_{\m\n}(\grad^\m\f)(\grad^\n\f)$; one can use the metric equation of motion to replace $R_{\m\n}$ in this term and obtain \eqref{eq:scalar} (see \cite{27_aron} for details).}
\begin{equation}
    \begin{aligned}\label{eq:scalar}
    &-\square\phi \Bigl(1 + \partial_X G_2 + 2 \partial_\phi G_3 - 2 X \partial^2_{\phi X}G_3\Bigr)- \partial_\phi G_2 + \partial_\f V\\
    &+2\,X(\partial^2_{\phi X}G_2 + \partial^2_{\phi\phi} G_3)  + (\partial^2_{XX} G_2 + \partial^2_{\phi X}G_3)(\nabla^{\mu}\phi)( \nabla^{\nu}\phi) \nabla_{\mu}\nabla_{\nu}\phi \\
    &+X\,\partial_X G_3\bigl(G_2 - V + X\,\br{2 + \partial_X G_2 + 4\, \partial_\phi G_3}\bigr)\\
    &+ X\,\bigl(\partial_X G_3\bigr)^2\big[X\,\square\phi + 2\,(\nabla^{\mu}\phi)(\nabla^{\nu}\phi)\nabla_{\mu}\nabla_{\nu}\phi\big] \\
    &+\partial^2_{XX}G_3(\nabla^{\mu}\phi)(\nabla^{\nu}\phi)\bigl[(\square\phi) \nabla_{\mu}\nabla_{\nu}\phi - (\nabla_{\mu}\nabla^{\rho}\phi) \nabla_{\rho}\nabla_{\nu}\phi \bigr] \\
    &- \partial_X G_3\Big[\bigl( \square\phi\bigr)^2 - (\nabla^{\mu}\nabla^{\nu}\phi) \nabla_{\mu}\nabla_{\nu}\phi\Big]  = 0\,.
    \end{aligned}
\end{equation}

We write down equations \eqref{eq:einstein} and \eqref{eq:scalar} in the usual 3+1 conformal decomposition and implement the CCZ4 form of the Einstein equations that is suitable for the numerical simulations. The equations that we have implemented in our code as well as the details of the numerical simulations are given in Appendix \ref{appendix:ccz4}. In the remainder of this Section, we describe the specific cubic Horndeski theories that we have studied, our initial data, the analysis of the hyperbolicity of the scalar equations and the weak field regime.  

\subsection{Cases explored}

The action \eqref{eq:action} comprises several well-known particular cases that have been extensively studied in other contexts, mostly cosmology (see \cite{43_AllHorndeskiCases,44_ModifiedGravityReview,Kobayashi:2019hrl}). For instance, \textit{quintessence}, which consists of a simple scalar field minimally coupled to GR; this model is obtained by setting $G_2=G_3=0$ in \eqref{eq:action}. On the other hand, models of \textit{k-essence} are obtained by setting $G_3=0$ in \eqref{eq:action}, with the common choice of $G_2(\f,X)=f(\f)g(X)$ for arbitrary functions $f$ and $g$ of their arguments. Finally,  \textit{kinetic gravity braiding} \cite{40_AnotherExampleOfHorndeskiInCosmology_G2andG3terms}, also referred as \textit{Cubic Galileons} \cite{31_horndeski_examples_1, 33_horndeski_examples_3}, are obtained from \eqref{eq:action} by choosing $G_3\neq0$; this class of models is often simplified to the shift symmetric case, corresponding to $G_3(\f,X)=g(X)$, for an arbitrary function $g$. Therefore, the subclass of Horndeski theories that we consider is very rich and has multiple applications to gravitational physics and cosmology. 

In our work we are not interested in a particular model but rather in exploring general features of the non-linear physics encoded in cubic Horndeski theories. From the point of view of EFT, one would expect \eqref{eq:action} to be valid when the $G_2$ and $G_3$ terms are suitably small, which corresponds to $X$ being small.    Therefore, one can consider Taylor-expanding some general (smooth) functions $G_2$ and $G_3$ for small $X$ and keep only the leading order terms. With this in mind,  we therefore focus on the simplest non-trivial functions $G_2$ and $G_3$:
\begin{equation}
    \begin{aligned}\label{eq:relevantGs}
        G_2(\f,X) &= g_2\, X^2,\\
        G_3(\f,X) &= g_3\, X,
    \end{aligned}
\end{equation}
where $g_2$ and $g_3$ are arbitrary coupling constants with dimensions of $\text{Length}^2$ that we can tune.  These or similar choices have been considered in the literature before, namely in models of dark energy \cite{32_horndeski_examples_2, 34_horndeski_examples_4, 35_horndeski_examples_5,36_horndeski_examples_6, 40_AnotherExampleOfHorndeskiInCosmology_G2andG3terms,41_AnotherExampleOfHorndeskiInCosmology_G2andG3terms2}, and in studies of the fate of the Universe in cosmological bounces or inflationary models \cite{20_G2andG3examples,23_bounce_withG4,36_horndeski_examples_6, 70_G2G3case_Inflation}, among others \cite{21_L3Horndeski, 91_Horndeski_sqrt_example}. As we noted in the introduction, from the point of view of EFT our choice for $G_2$ in \eqref{eq:relevantGs} corresponds to the most general scalar term that can be added to the action up to four derivatives \cite{Weinberg:2008hq}.

\subsection{Initial data}
\label{sec:initial_data}

For the present analysis, motivated by the objective of studying gravitational collapse, we choose a family of initial data for the  scalar field $(\f,\P)$ modelling a spherically symmetric bubble centred at $\ve{c}$:
\begin{equation}\label{eq:phiData}
    \phi(t,\ve{x})\bigg\rvert_{t=0} = A \br{\frac{r^2}{r_0^2+2\,\omega^2}}e^{-\frac{1}{2}\br{\tfrac{r-r_0}{\omega}}^2},
\end{equation}
where $\ve{r}=\ve{x}-\ve{c}$ and $r=||\ve{r}||_2$ with the Euclidean 2-norm. 
Notice that the class of theories in \eqref{eq:relevantGs} have a reflection symmetry  $\f\to-\f\,,\,g_3\to -g_3$ and hence, we can choose $A>0$ without loss of generality. Regarding the scalar momentum, assuming an approximately Minkowski initial background, we choose an ingoing wave pulse:
\begin{align}\label{eq:ini_mom}
    \Pi(t,\ve{x})\bigg\rvert_{t=0} = \frac{1}{r}\partial_r\br{r\phi}\bigg\rvert_{t=0}\,.
\end{align}

To explore the relevant phenomenology of these theories, we have studied many different scenarios. Using a full 3D code, we were able to verify that all the features hereafter described are not a peculiarity of spherical symmetry, and also occur when the symmetry is broken, without any seemingly interesting new features emerging. However, we have not attempted to carry out a thorough analysis of non-spherically symmetric scalar field collapse.  Hence, in the following we only present the results for the spherically symmetric case. 

With the choices \eqref{eq:phiData} and \eqref{eq:ini_mom} for the initial scalar profile and momentum, we obtain the initial data for the metric by solving the Einstein constraints using the conformal transverse-traceless decomposition \cite{alcubierre,shapiro}. We choose a conformally flat initial metric and vanishing trace and transverse-traceless part of the extrinsic curvature. Hence, we solve for the conformal factor of the spatial metric and three leftover degrees of freedom of the traceless part of the extrinsic curvature (which reduce to one in spherical symmetry).

To get some intuition about how the modifications of GR affect our initial data, we can expand the initial ADM mass for small amplitudes and couplings around a Minkowski background. We find,
\begin{align}
    M_{\text{ADM}} \approx&~ \frac{\bar{\rho}}{16\,\pi}\,\Bigg\{1+\frac{13}{2}\br{\frac{\omega}{r_0}}^2 + \mathcal{O}\bigg(\br{\frac{\omega}{r_0}}^4\bigg) \label{eq:rho0} \\
    &\hspace{0.5cm}+\bigg[\br{m^2\omega^2}+\br{\frac{5}{4\sqrt{2}}}\br{\frac{g_2A^2}{r_0^2}}+\bigg(\frac{28}{9}\sqrt{\frac{2}{3}}\bigg)\br{\frac{g_3A\omega^2}{r_0^4}}\bigg]\bigg[1+\mathcal{O}\bigg(\br{\frac{\omega}{r_0}}^2\bigg)\bigg]\Bigg\}\,, \nonumber
\end{align}
where $\bar{\rho}=\frac{\sqrt{\p}}{8}\frac{A^2r_0^2}{\omega}$ and we have included the contribution of a mass term in the scalar potential $V(\f)=\frac{1}{2}m^2\f^2$. From \eqref{eq:rho0} we see that for our initial data, the strength of the modifications of GR due to the Horndeski terms is measured by the dimensionless couplings:
\begin{equation}
\eta_2 = \frac{g_2A^2}{r_0^2}\,, \label{eq:eta2}
\end{equation}
and,
\begin{equation}
\eta_3 = \frac{g_3A\omega^2}{r_0^4}\,, \label{eq:eta3}
\end{equation}
respectively. These dimensionless couplings play an important role in the future development of the initial data and determine the weakly coupled regime of these theories. 

\begin{figure}[t!]
\centering
\includegraphics[width=.8\textwidth]{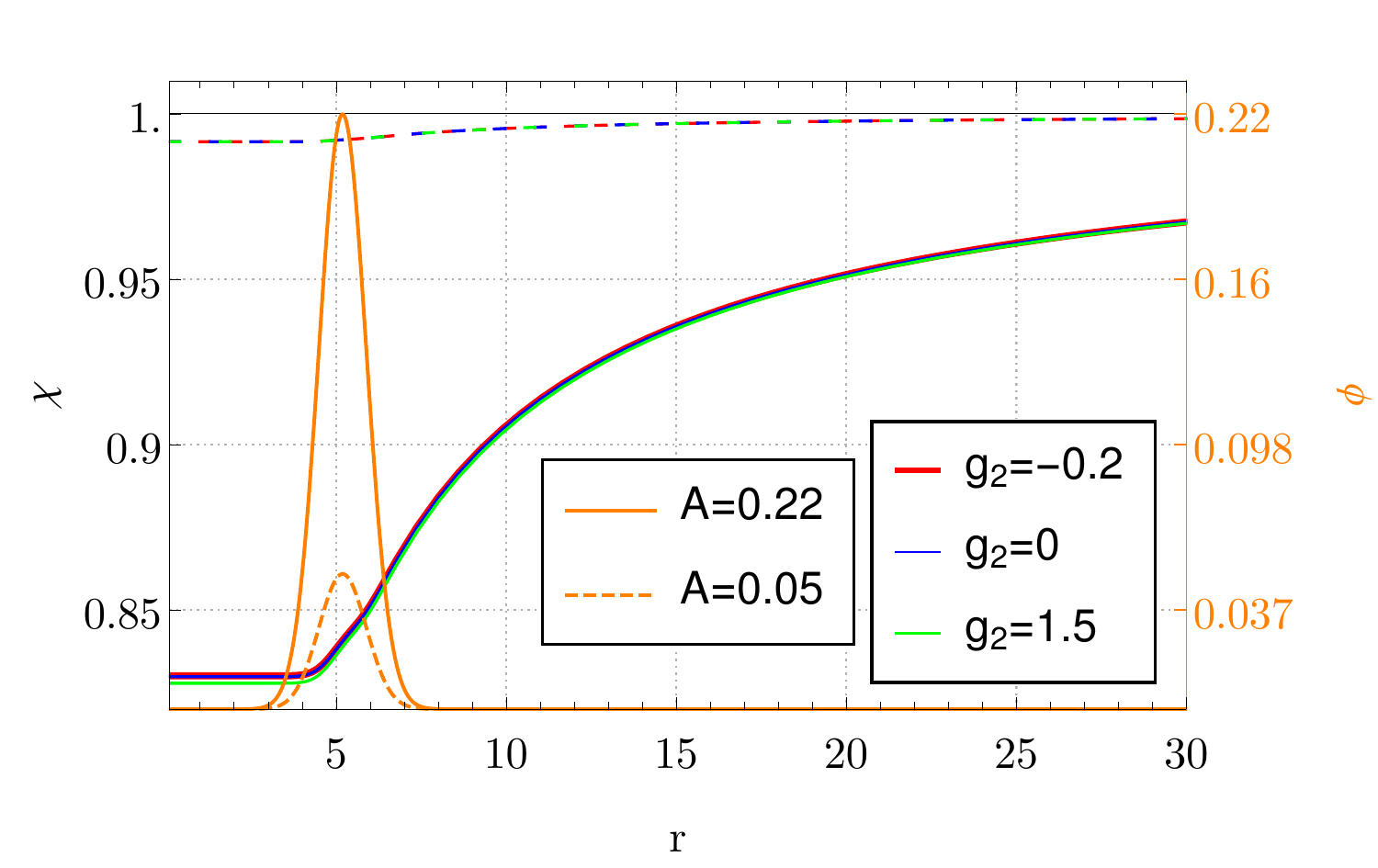}
\vspace{-3mm}
\caption{Initial conformal factor $\chi$ (left $y$-axis) for a given scalar field profile (in orange, right $y$-axis) with ingoing momentum, for different choices of $g_2$. The scalar profile in this figure corresponds to Cases 1 (dashed line) and 3 (solid line) of Section \ref{numerics}; the shown values of coupling $g_2$ correspond to the GR case ($g_2=0$) and the cases presented in Fig. \ref{fig:strong_g2_pos_det_h00} ($g_2=1.5$) and Fig. \ref{fig:strong_g2_neg_det_h00} ($g_2=-0.2$).}
\label{fig:initial_data}
\end{figure}

In Fig. \ref{fig:initial_data} we show the initial conformal factor $\chi$ and scalar profile $\phi$ for some representative cases.  From this figure we see that even for relatively large amplitudes within the range that we have considered, the conformal factor has a very small dependence on the Horndeski couplings. For the specific case of $A=0.22$, the difference between $g_2=1.5$ and GR at $r=0$ is $0.2\%$, which is in accordance with the fact that for this case the dimensionless coupling is small $(\eta_2\sim 3\times 10^{-3})$. One can also notice that for sufficiently small amplitude, the conformal factor is almost 1 for any reasonable value of $g_2$.

\subsection{Effective metric and characteristic speeds}
\label{sec:effmetric}

To identify the regime of validity of EFT, we need to first determine the character of the equations of motion for the scalar field \eqref{eq:scalar} and the conditions under which they are hyperbolic. To do so, we consider the principal part of the scalar equation \eqref{eq:scalar}, which is a wave equation governed by an effective metric \cite{40_AnotherExampleOfHorndeskiInCosmology_G2andG3terms}:
\begin{equation}
\begin{aligned}
    h^{\mu\nu} =&~~~g^{\mu\nu}\,\Big[1 + \partial_X G_2 + 2\, \partial_\phi G_3 + 2\,\partial_X G_3 \,\Box\phi - X^2 \bigl(\partial_X G_3\bigr)^2 \\
    &\hspace{1.2cm}- \partial^2_{XX} G_3 (\grad^\rho\phi)(\grad^\sigma\phi)\grad_\rho\grad_\sigma\phi - 2\,X\,\partial^2_{\phi X} G_3\Big]\\
    & -(\grad^\mu\phi)(\grad^\nu\phi)\sbr{2\,X\bigl(\partial_X G_3\bigr)^2 + \partial^2_{XX} G_2 + \partial^2_{XX} G_3\,\Box\phi + 2\,\partial^2_{\phi X} G_3}\\
    & +~2\,\partial^2_{XX}G_3(\grad^\rho\phi)(\grad^{(\mu}\phi)\grad_\rho\grad^{\nu)}\phi - 2\,\partial_X G_3 \grad^\mu\grad^\nu\phi\,.
    \label{eq:eff_metric_up}
\end{aligned}
\end{equation}
The eigenvalues of $h^{\mu\nu}$ determine the character of the equation: if the product of the eigenvalues is negative  then equation is hyperbolic; if the product is positive then the equation is elliptic, and if it is zero the equation is parabolic.  
For the specific cases considered in this paper, see \eqref{eq:relevantGs}, the effective metric is given by
\begin{align}
	h^{\mu\nu} =&~ g^{\mu\nu}\br{1 + g_2\, X} - 2g_2\,(\grad^\mu\phi)(\grad^\nu\phi)\,,\\
    h^{\mu\nu} =&~ g^{\mu\nu}\,\left(1 + 2g_3\, \,\Box\phi - g_3^2\,X^2\right) - 2g_3^2\,X\,(\grad^\mu\phi)(\grad^\nu\phi) - 2g_3\, \grad^\mu\grad^\nu\phi\,,
\end{align}
respectively.

Having a well-posed initial value problem is the minimum requirement that we should demand on any classical theory; therefore, the breakdown of hyperbolicity of the scalar equation in this case can be associated to the breakdown of the theory itself.  As \cite{125_caustics_in_Horndeski,38_horndeski_new_may_example_luis} noted, the fact that the effective metric \eqref{eq:eff_metric_up} depends on the scalar field itself and its gradients implies that shocks can generically form from smooth initial data; at that point uniqueness is lost which in turn could lead to a loss of well-posedness.  Therefore, the local character of the scalar equation is a useful proxy to establish the regime of validity of the theory and to measure the size of the non-linearities and deviations from GR \cite{38_horndeski_new_may_example_luis,25_pretorius_horndeski_19,39_Pretorius_EdGB_elliptic,73_pretorius_EdGB_elliptic_inside_BHs}. We will come back to this point below.  

When considering spacetimes containing black holes, the evolution of the spatial slices in puncture gauge is such that the determinant of the inverse spacetime metric goes to zero near the puncture, i.e., $\det(g^{\m\n})=-\frac{\chi^3}{\a^2}\to0$ (see Appendix \ref{appendix:ccz4}). Consequently the same happens for the effective metric \eqref{eq:eff_metric_up}. To distinguish this gauge effect from an actual breakdown of the hyperbolicity of the scalar equation, we note that $h^{\mu\nu} = g^{\mu\rho}h^{\nu}_{~\rho}$ and therefore:
\begin{equation}
    \text{det}\br{h^{\mu\nu}} = \text{det}\br{h^\nu_{~\rho}}\,\text{det}\br{g^{\mu\rho}} = -\frac{\chi^3}{\alpha^2}\,\text{det}\br{h^\mu_{~\nu}},
    \label{eq:dethuu}
\end{equation}
with $\text{det}\br{h^\mu_{~\nu}}=1$ in GR. Clearly, deviations of this quantity from 1 encode the dynamics of the Horndeski theories and hence we will focus our attention on $\text{det}\br{h^\mu_{~\nu}}$.

The characteristic speeds, also called front velocities, are important since they correspond to the local speed of propagation of the scalar modes and hence they tell us about the effective causal cone that the scalar field ``sees".\footnote{Recall that the characteristic speeds do not coincide in general with the phase or group velocity, which do not have a direct relation with the causal structure.} The characteristics are given by the zeros of the characteristic polynomial which, for the scalar field equation, is
\begin{equation}
    Q(x,\xi) = h^{\mu\nu}\xi_\mu\xi_\nu=0\,,
\end{equation}
for some covector $\xi_\m$ that defines the characteristic surface. Physically this corresponds to considering the high frequency and small amplitude limit of a wave with wave vector $\xi_\mu$. To calculate the propagation speeds without symmetry assumptions, we specify a direction of propagation, $n_i$ suitably normalised $n_i n_j \d^{ij}=1$, where $\delta_{ij}$ is the Euclidean 3D metric (since the space is locally flat). Then, the speed of propagation in the $n^i$ direction is:
\begin{gather}
    h^{00}v^2 + 2h^{0i}n_i ~v + h^{ij}n_i n_j = 0 \quad \Rightarrow\quad
    v_{\pm} = \frac{-h^{0i}n_i \mp \sqrt{\br{h^{0i}n_i}^2 - h^{00}h^{ij}n_i n_j}}{h^{00}}\,.
\end{gather}
In spherical symmetry one can naturally use a radial vector for the direction of propagation, $n^i=\{\tfrac{x}{r},\tfrac{y}{r},\tfrac{z}{r}\}$, which gives \cite{Babichev:2018uiw,38_horndeski_new_may_example_luis},
\begin{equation}
v_{\pm}=\frac{-h^{0r}\mp\sqrt{\br{h^{0r}}^2-h^{00}h^{rr}}}{h^{00}}\,.
\end{equation}
In our conventions, $v_+$ and $v_-$ correspond to the ingoing and outgoing modes of the scalar field respectively and they are normalised so that they tend to $+1$ and $-1$ at infinity. When $v_-\geq 0$ and $v_+\geq 0$ in a certain region, scalar modes cannot reach asymptotic observers; the boundary $v_-=0$ of this region is the sound horizon \cite{Babichev:2006vx,67_G2Horndeski_gtt_to_zero}.  The characteristic speeds of propagation with respect to proper time are obtained rescaling $v_\pm$ by a factor of $1/\a$:
\begin{gather}\label{eq:vproper}
v^{\text{proper}}_\pm = \frac{v_{\pm}}{\a} \,.
\end{gather}
Since in our working gauge the lapse $\alpha$ is strictly positive everywhere except at the `puncture',\footnote{In fact, we effectively excise a region inside the AH that contains the `puncture'.} $v_\pm$ and $v^{\text{proper}}_\pm$ carry the same practical information; in particular, the sound horizons will be located at the same place.

As discussed in \cite{39_Pretorius_EdGB_elliptic,38_horndeski_new_may_example_luis}, the equations can change character from hyperbolic to parabolic and elliptic in a manner which is qualitatively similar to what happens in the two standard equations of mixed type, namely the Tricomi equation,
\begin{equation}
    \partial_y^2u(x,y)+y\,\partial_x^2 u(x,y)=0\,,
\end{equation}
and the Keldysh equation,
\begin{equation}
    \partial_y^2u(x,y)+\frac{1}{y}\,\partial_x^2 u(x,y)=0\,,
\end{equation}
Both equations are hyperbolic for $y<0$ and they change character at the transition line $y=0$. Related to this change of character are the appearence of ghosts, gradient instabilities and formation of caustics \cite{65_ghostsCubicHorndeski,72_GhostAndLaplacianInstability_Horndeski, 125_caustics_in_Horndeski}. 

For a hyperbolic equation, the characteristic speeds should be real and finite. In the case of the Tricomi equation, the characteristic speeds go to zero at  $y=0$ where the equation becomes parabolic, while in the Keldysh equation the characteristic speeds diverge at $y=0$. If the characteristic speeds of both the ingoing and outgoing modes vanish, then the evolution freezes. This can happen because of the choice of gauge; for instance, in coordinates that are not horizon penetrating, the lapse asymptotically goes to zero at the horizon, effectively resulting in zero characteristic speeds.  However, in this case the freezing of the evolution is a consequence of the gauge choice and it does not correspond to a breakdown of EFT. Therefore, in the case of a Tricomi-type-of transition, we also need to check that the deviations from GR are suitably large to conclude that the loss of hyperbolicity corresponds to a breakdown of the theory.  On the other hand, a Keldysh-type-of transition involves diverging characteristic speeds,\footnote{At least in some direction in full 3D space, which is non-trivial to determine without spherical symmetry.} which will typically signal a breakdown of EFT. This case is more difficult to handle numerically since  one is forced to take prohibitively small time steps.\footnote{In fact, the degree of regularity of the solutions of these equations typically differs, with solutions of the Tricomi equation enjoying higher regularity \cite{39_Pretorius_EdGB_elliptic}.} Note from \eqref{eq:eff_metric_up} that $h^{00}$ has a factor of $-1/\alpha^2$ coming from $g^{00}$, 
and hence the deviations from GR are measured by $-\alpha^2 h^{00}$. Therefore, the Keldysh-type-of transition without symmetry assumptions is signalled by $-\a^2 h^{00}\to0$, which implies that the $t=\textrm{const}.$ hypersurface being evolved is no longer spacelike with respect to the scalar effective metric \cite{119_Horndeski_ClosedCausalCurves_InitialDataValidity}. We associate this breakdown of the evolution to a Keldysh-type-of transition since the characteristic speeds diverge.  However, strictly speaking, at this point the equation may not have changed character yet but the two effects go hand in hand.\footnote{We would like to thank Luis Lehner for discussions on these issues.}  In practice, since we always start from a hyperbolic equation, by continuity the breakdown of the evolution happens because of a Tricomi or a Keldsyh-type of transtion. Either of those occurs before an elliptic region forms. For this reason, in our simulations we do not observe the appearance of elliptic regions and hence we will not dwell on this case any further. We discuss in detail the different types of transitions in the $G_2\neq0$ and $G_3\neq0$ cases in the next subsection.

The previous discussion only relates to the existence of a well-posed initial value problem but it does not fully address the issue of whether the theory under consideration makes sense as a truncated EFT \cite{94_EFTs}. We now turn to this point. As mentioned in \cite{27_aron}, local well-posedness is only guaranteed in the weak field regime, meaning that the Horndeski terms are small compared to GR ones. One possible weak field condition that compares the size of the Horndeski terms versus GR is:
\begin{equation}
\begin{aligned}\label{eq:WFC_1}
    &\left|\partial^k_X\partial^l_\phi G_2\right| \ll L^{2k-2} ~~~~~~~~~~~~~~~ k=0,1,2;~l=0,1;\\
    &\left|\partial^k_X\partial^l_\phi G_3\right| \ll L^{2k} ~~~~~~~~~~~~~~~~~~~~~~~~~~~ k,l=0,1,2.
\end{aligned}
\end{equation}
where $L$ is a length scale estimate for the system: $L^{-1}=\text{max}\{\left|R_{\alpha\beta\mu\nu}\right|^{\tfrac{1}{2}},\left|\grad_\mu\phi\right|,\left|\grad_\mu\grad_\nu\phi\right|^{\tfrac{1}{2}}\}$ in all orthonormal bases. For the cases \eqref{eq:relevantGs}, this is explicitly:
\begin{gather}\label{eq:WFC_2}
     |g_2 \, L^{-2}| \ll 1\,,  \quad\quad  |g_3\, L^{-2}| \ll 1\,.
\end{gather}
In order for the Horndeski theories under consideration \eqref{eq:relevantGs} to be in the regime of validiy of EFT, in this paper we require that the evolution equation of the scalar field is hyperbolic \textit{and} that \eqref{eq:WFC_2} is satisfied. These two conditions ought to be imposed on and outside black hole horizons, should there be any in the spacetime. 

\subsubsection{Case of $G_2\neq 0$, $G_3=0$}
\label{effective_metric_G2}

To monitor the character of the scalar equation, we  compute the determinant of the scalar effective metric. Even though it is possible to find an analytic expression for the full determinant (using Cayley–Hamilton’s theorem and Newton’s identities), for simplicity we consider the $G_2\neq 0$, $G_3=0$ and the $G_3\neq 0$, $G_2=0$ cases separately. 

As explained in the discussion surrounding eq. \eqref{eq:dethuu}, we only need to consider the determinant of the effective metric with one index up and one index down, which is significantly simpler. For the $G_2\neq 0$ case, we have
\begin{gather}
    h^{\mu}_{~\nu} = \delta^{\mu}_{~\nu}\br{1 + \partial_X G_2} - (\grad^\mu\phi)(\grad_\nu\phi)~\partial^2_{XX} G_2.
\end{gather}
Realising that, up to scalars, this metric is the identity plus the tensor product of two vectors, one can use the Weinstein–Aronszajn identity to calculate the determinant of the full 4D metric without assuming any symmetries. We find:
\begin{equation}
    \begin{aligned}\label{eq:det_h00G2}
    \text{det}\br{h^\mu_{~\nu}}=&~\br{1+\partial_X G_2}^3\br{1 + \partial_X G_2 + 2\,X\partial^2_{XX} G_2}\\
    =&~\br{1+2\,g_2X}^3\br{1+6\,g_2\,X}\,,
    \end{aligned}
\end{equation}
where in the last line we have used that $G_2=g_2\,X^2$. We can compute the eigenvalues and eigenvectors $v^\nu$ by noting that 
\begin{equation}
    h^\mu_{~\nu}v^\nu = v^\mu\br{1 + \partial_X G_2} - \br{\partial^2_{XX}G_2v^\nu\grad_\nu\phi }\grad^\mu\phi\,,
\end{equation}
so we conclude that $\grad^\mu\phi$ is an eigenvector with eigenvalue $\br{1 + \partial_X G_2 + 2\,X\partial^2_{XX} G_2}$. The other three eigenvectors are orthogonal to the 4-vector $\nabla^\m\f$ and have degenerate eigenvalues equal to $\br{1+\partial_X G_2}$, in accordance to \eqref{eq:det_h00G2}.

To monitor a Keldysh-type-of transition, we have to compute $-\a^2h^{00}$. For the $G_2\neq 0$, $G_3=0$ case, this is given by,
\begin{equation}
\begin{aligned}
    -\a^2 h^{00}&=1+\partial_X G_2 + \Pi^2~\partial^2_{XX}G_2 \\
        &=1+6\,g_2X+2\,g_2\,\Pi^i\Pi_i\,,\label{eq:h00G2}
\end{aligned}
\end{equation}
where $\P=n^\m\grad_\m\f$ is the scalar momentum, and in the last line we have used that $\Pi^2 = 2\,X + \Pi^i\Pi_i$, with $\P_i=D_i\f$ (see Appendix \ref{appendix:ccz4}). All in all, for $G_2$ as in \eqref{eq:relevantGs}, the two quantities that inform us about the breakdown of the initial value problem for the scalar equation are \eqref{eq:det_h00G2} and \eqref{eq:h00G2}.
The scalar equation is hyperbolic as long as these two quantities are non-negative.\footnote{Note that we have pulled out a minus sign in \eqref{eq:dethuu} so $\det\br{h^\mu_{~\nu}}>0$ corresponds to $h^{\m\n}$ having one negative eigenvalue and three positive ones, as it should for a hyperbolic equation.} These are the same conditions found  in \cite{119_Horndeski_ClosedCausalCurves_InitialDataValidity}, and  it is evident that if the weak field conditions \eqref{eq:WFC_2} are satisfied then the scalar equation is hyperbolic.  

\begin{figure}[t!]
\centering
\includegraphics[width=.6\textwidth]{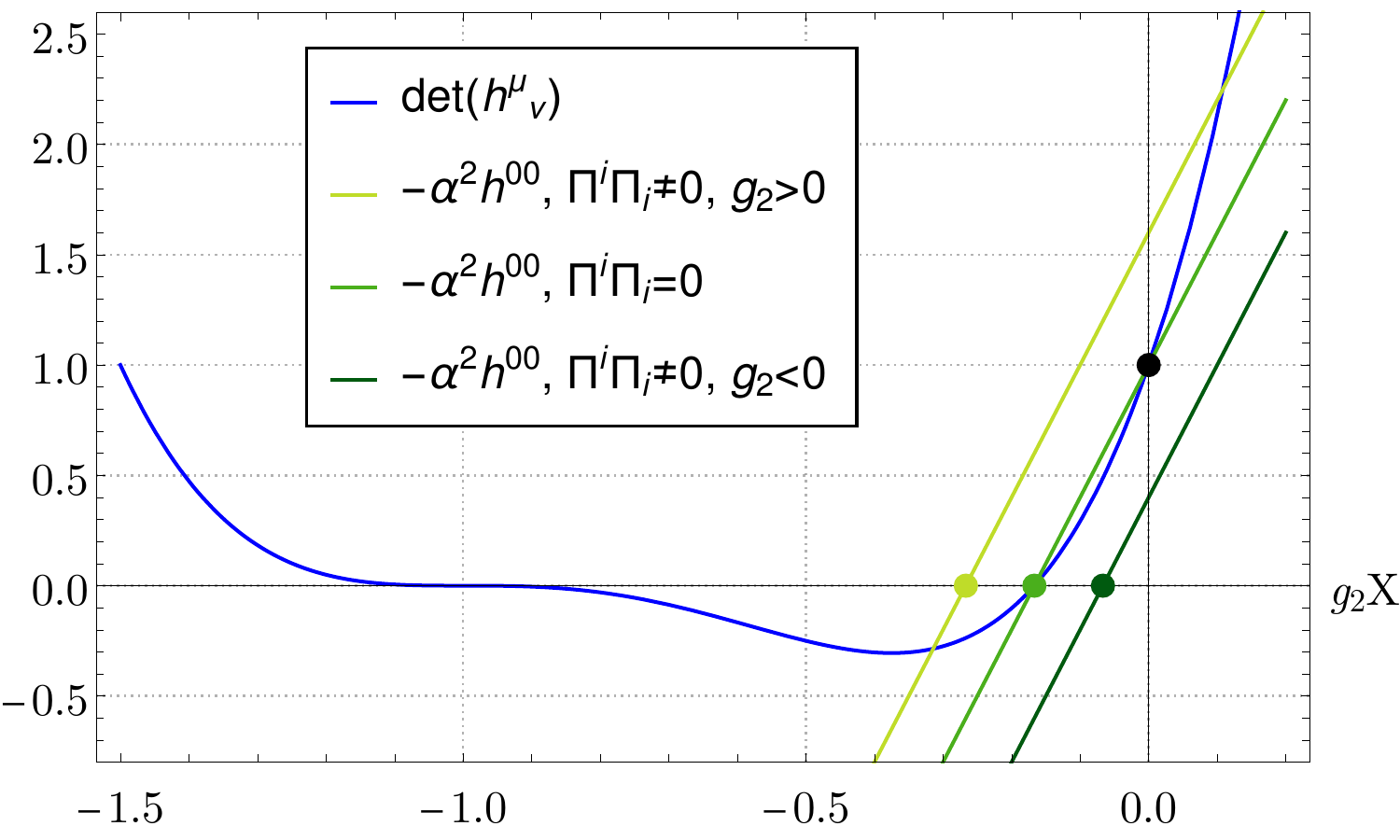}
\vspace{-3mm}
\caption{Sketch of the possible changes of character of the scalar equation in the $G_2\neq 0$ theory depending on the sign of the coupling constant $g_2$. For $g_2>0$, $\det(h^\m_{~\n})$ (blue curve) vanishes before $-\a^2h^{00}$ does (light green line), leading to a Tricomi-type-of transition. On the other hand, for $g_2<0$, $-\a^2h^{00}$ will vanish first (dark green line), leading to a Keldysh-type-of transition. In this case, the time coordinate $t$ is no longer a global time function and the scalar equation cannot be evolved further in this gauge. }
\label{fig:plot_g2X_det_and_h00}
\end{figure}

For a non-constant scalar profile $\f$, $\P_i\P^i>0$ but $X$ can be either positive or negative, depending on the balance between scalar gradients and  momentum. In a dynamical evolution, both can become large. As this happens, $g_2\,X$ can decrease to make either  \eqref{eq:det_h00G2} or \eqref{eq:h00G2} zero, see Fig. \ref{fig:plot_g2X_det_and_h00}. If $g_2>0$, the fact that $\P_i\P^i>0$ implies that $\text{det}\br{h^\mu_{~\nu}}$ will reach zero before $-\a^2h^{00}$, and the equation will become parabolic on a co-dimension one surface, where at least one of the  characteristic speeds goes to zero while the others remain bounded. This will correspond to a Tricomi-type-of transition.  On the other hand, if $g_2<0$ the opposite is true and $-\a^2h^{00}$ may become zero before $\text{det}\br{h^\mu_{~\nu}}$ does, leading to infinite speeds of propagation and a very abrupt termination of the evolution associated to a Keldysh-type-of transition. 
Both behaviours were identified  in \cite{38_horndeski_new_may_example_luis}.\footnote{Reference \cite{38_horndeski_new_may_example_luis} uses a coupling $g$ with the opposite sign as our $g_2$.}

The changes of character described in the previous paragraph can only occur if $|g_2\,X|$ is suitably large and hence outside the weak field regime. While generically one can expect that weak data eventually enters the strong field regime,  one question that we need to address  is whether or not the region where EFT breaks down can be hidden inside a black hole. If the answer is positive, then one can hope that classical observers at infinity will be protected from any potential pathologies that arise in the scalar equations and EFT will retain its predictive power. The technical details on how we have dealt with the loss of hyperbolicity and the violations of the weak field condition \eqref{eq:WFC_2} inside black holes are given in Appendix \ref{appendix:excision}.

\subsubsection{Case of $G_3\neq 0$, $G_2=0$}
\label{effective_metric_G3}

In equation \eqref{eq:det_full_g3} of Appendix \ref{appendix:det_eff_metric} we present the full analytic form of the determinant of the scalar effective metric in the $G_3\neq 0$, $G_2=0$ case. For clarity, in this subsection we analyse \eqref{eq:det_full_g3} for small $g_3$, which is the relevant limit in the weak field regime. 

To obtain the expansion of \eqref{eq:det_full_g3} for small $g_3$, we use the scalar equation of motion (several times if necessary) to replace $\Box\f$ in \eqref{eq:det_full_g3} by $V'(\f)$ and terms which are higher order in $g_3$, in the spirit of order reducing schemes. We then obtain, up to second order:
\begin{equation}
\begin{aligned}
    \text{det}\br{h^\m_{~\n}} =&~ 1 + 6\,g_3\,V'(\f) + g_3^2\sbr{-6\,V(\f)\,X + 8\,V'(\f)^2 + 12\,X^2 + 4\br{\grad_\m\grad_\n\f}\br{\grad^\m\grad^\n\f}} \\
    &+\mathcal{O}\br{g_3^3}
\end{aligned}
\end{equation}
Similarly, we find:
\begin{equation}
    -\a^2 h^{00} = 1+2\,\t g_3 - g_3^2 \br{X^2 - 2\,\P^2 X}+\mathcal{O}(g_3^3)\,,
\end{equation}
where $\t=K\,\P+D^i\P_i$ is independent of $g_3$, see \eqref{eq:auxvars}. 

Let us focus on the case of zero scalar potential, $V\br{\f}=0$, which is the relevant one for this paper. In this case, the correction to GR in $\det(h^\m_{~\n})$ comes at $\mathcal O(g_3^2)$, while in $ -\a^2 h^{00}$ it comes at leading order. Because $\tau$ does not have a definite sign, then regardless of the sign of $g_3$, there will be regions in spacetime where $ -\a^2 h^{00}$ will vanish before $\text{det}\br{h^\m_{~\n}}$ does, resulting in a Keldysh-type-of transition. This should be the generic behaviour in the $V\br{\f}=0$ case for the $G_3\neq 0$, $G_2=0$ theory, and it is  indeed what we observe in our numerical simulations, see Section \ref{sec:G3_neq_0}. The picture changes for $V(\f)\neq 0$; then  $\text{det}\br{h^\m_{~\n}}$ receives a contribution to leading order in $g_3$ and the type of transition will depend on the details of the scalar potential and the initial data.

\section{Numerical results}\label{numerics}
In this section we present the results of our numerical simulations of the gravitational collapse of a single massless scalar bubble with intial data as in Section \ref{sec:initial_data}. In all our simulations we keep the radius $r_0$ and the width $\omega$ of the initial Gaussian profile \eqref{eq:phiData} fixed, and we vary both the amplitude $A$ and Horndeski coupling $g_2$ or $g_3$. The reason is that varying $r_0$ and $\omega$ leads to similar results and varying $A$ alone makes the analysis simpler. We choose $r_0=5$ and $\w=\sqrt{0.5}$, which set the length scale in our simulations. 

\begin{figure}[t!]
\centering
\includegraphics[width=\textwidth]{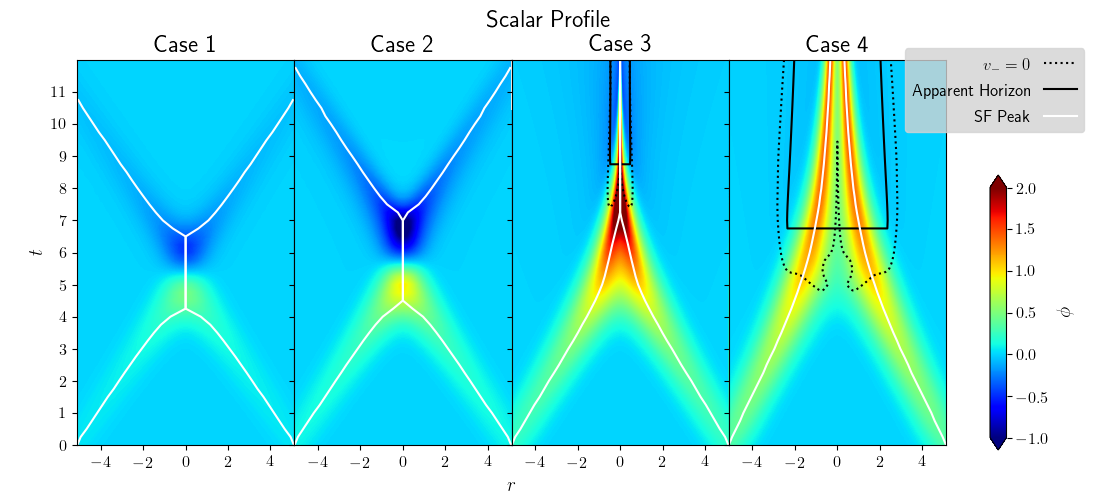}
\vspace{-5mm}
\caption{Evolution of the profile of a (massless) scalar field in GR for the Cases 1--4.}
\label{fig:GR_scalar_profile}
\end{figure}

Since we consider spherically symmetric scalar field collapse (even though we do not assume spherical symmetry in our simulations), there are essentially two relevant regimes depending on whether the initial data disperses to infinity (small data) or it collapses into a black hole (large data). We consider four representative values of the initial amplitude $A$, so that we can probe the regimes far and close to critical collapse for both small and large data:\footnote{The endpoints in Cases 1--4 below are obtained by evolving our initial data turning off \textit{all} Horndeski terms, see Fig. \ref{fig:GR_scalar_profile}. For large enough couplings, the equations may break down before the scalar field has either dispersed or collapsed into a black hole.}
\begin{itemize}
    \item \textbf{Case 1:} $A=0.05$ -- dispersion far from the critical regime, with initial ADM mass of $\approx0.022$.
    \item \textbf{Case 2:} $A=0.10$ -- dispersion closer to critical regime, with initial ADM mass of $\approx0.1$.
    \item \textbf{Case 3:} $A=0.22$ -- collapse into small black hole with initial ADM mass of $\approx0.5$.
    \item \textbf{Case 4:} $A=0.33$ -- collapse into a larger black hole with initial ADM mass of $\approx1.6$.
\end{itemize}
For each of these cases, we vary the Horndesky couplings ($g_2$ or $g_3$) while ensuring that the initial value problem is well-posed. We then  evolve the initial data  by solving the coupled equations of motion \eqref{eq:einstein}--\eqref{eq:scalar} numerically, and we monitor both the hyperbolicity of the scalar equation and the weak field conditions \eqref{eq:WFC_2}. In this way we can identify the regime of validity of the EFT for both small and large data.  We shall refer to the different cases as ``weakly" or ``strongly" coupled depending on the whether the hyperbolicity of the scalar equation breaks down at some point during the evolution; this breakdown is associated to the weak field conditions \eqref{eq:WFC_2} becoming large compared to the dimensionless couplings \eqref{eq:eta2} and \eqref{eq:eta3}.  The  evolution of the scalar field in GR (i.e., $g_2=g_3=0$) for the Cases 1--4 is shown in Fig. \ref{fig:GR_scalar_profile}.

It is  worth emphasising that Cases 2 and 3 above do not exhibit Choptuik's critical behaviour, as the amplitude $A$ is purposely chosen to be sufficiently `far' from the critical amplitude $A_*\approx 0.13\pm0.01$. The reason is that Choptuik's critical solution is a naked singularity and EFT will necessarily break down close to it. Indeed, zooming in near the black band in  Fig. \ref{fig:results_eta2} and  Fig. \ref{fig:results_eta3} would show that the coupling constants have to be tuned down to maintain the hyperbolicity of the scalar equation as we approach the critical regime from both sides. In addition, the weak field conditions \eqref{eq:WFC_2} become large the closer we get to the critical solution, as expected.

Since the $G_2\neq 0$ and $G_3\neq 0$ theories do not exhibit significant qualitative differences in terms of the dynamics of collapse of the scalar field, in the next subsection we will focus the discussion on the $G_2\neq 0$ theory considering different values and signs of the coupling constant $g_2$. In subsection \ref{sec:G3_neq_0} will only highlight the main differences in the $G_3\neq 0$ case.

\subsection{$G_2$ theories}
\label{sec:G2_neq_0}

In the following subsections we will discuss gravitational collapse in Horndeski theories with $G_2 = g_2\,X^2$ for different values of the coupling constant $g_2$.  For our scalar field initial data, during collapse a positive and negative peak in $X$ form; these peaks grow as the evolution progresses and the scalar shell approaches the origin. After reaching the origin, they bounce back and smaller peaks of opposite signs form,  eventually resulting in the formation of a black hole or dispersion to infinity. See Fig. \ref{fig:GR_scalar_profile} for the evolution of the scalar field profile in GR; in the Horndeski theories it is qualitatively similar. With an initial ingoing momentum, as in our initial data, momentum dominates over spatial gradients and the positive peak will be much larger in amplitude than the negative or the subsequent peaks that form after the bounce. Considering the expressions for $\det(h^\m_{~\n})$ and $-\a^2\,h^{00}$ in \eqref{eq:det_h00G2} and \eqref{eq:h00G2} for the $G_2\neq 0$ theory, this implies that generically a negative $g_2$ will lead to a breakdown of the hyperbolicity of the equations for a significantly smaller $|g_2|$ and it will happen sooner than for a positive $g_2$. Furthermore, as described in Section \ref{effective_metric_G2}, for $g_2<0$ the change of character will be of the Keldysh type while for $g_2>0$ it will be of the Tricomi type. 

\subsubsection{Weak coupling}
\label{sec:g2_pos_weak_coupling}

We first consider the case of a small and positive coupling constant $g_2$; we choose $g_2=0.005$ as a representative example.  This is a case of a theory that remains in the regime of validity of EFT throughout the whole evolution, both for small and large initial data. For this choice of parameters, the maximum of the weak field condition \eqref{eq:WFC_2} is small everywhere on and outside horizons (if they form) at all times. 

\begin{figure}[t!]
\centering
\includegraphics[width=\textwidth]{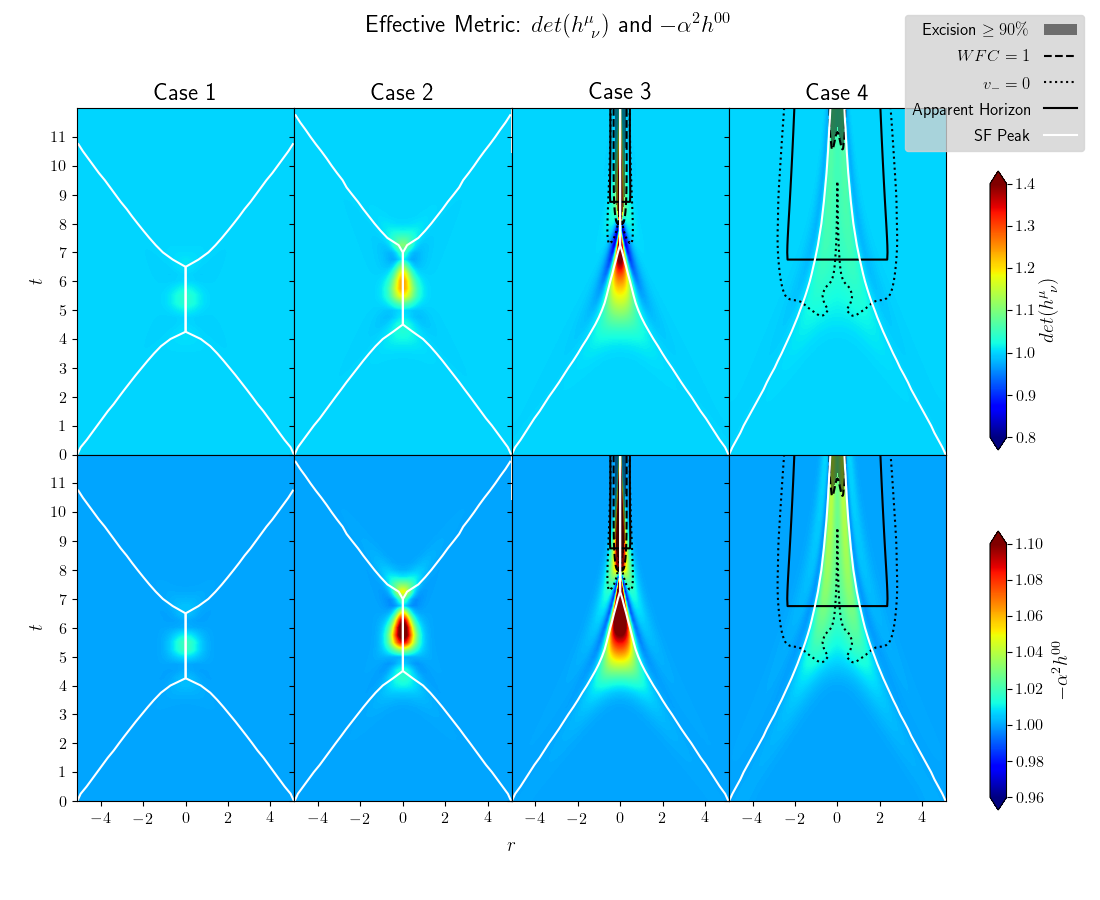}
\vspace{-5mm}
\caption{$\det (h^\m_{~\n})$ (\textit{top}) and $-\a^2\, h^{00}$ (\textit{bottom}) for $g_2=0.005$. The corresponding values of $\eta_2$ are, from left to right, $5\times 10^{-7},\,2\times 10^{-6},\,9.7\times 10^{-6},\,2.2\times 10^{-5}$ respectively.}
\label{fig:weak_g2_pos_det_h00}
\end{figure}
%

%
\begin{figure}[t!]
\centering
\includegraphics[width=\textwidth]{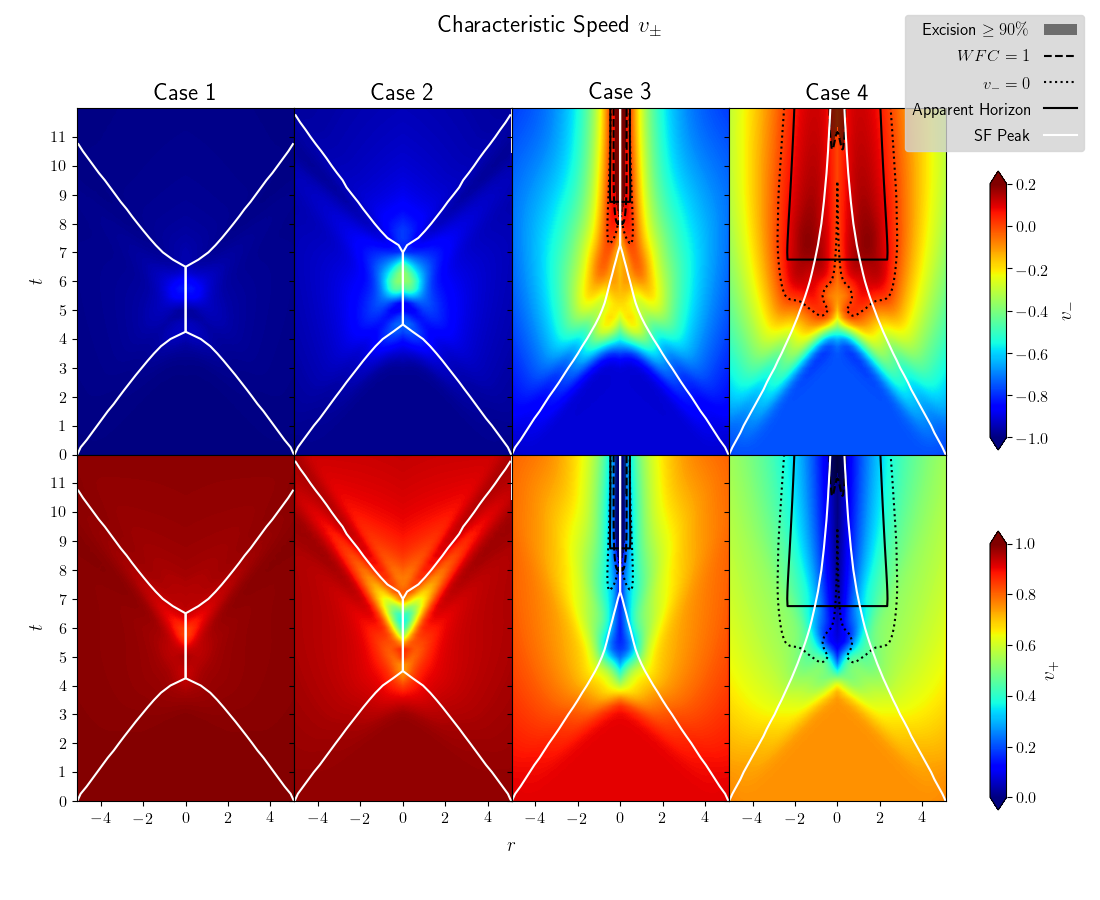}
\vspace{-5mm}
\caption{Outgoing (\textit{top}) and ingoing (\textit{bottom}) scalar characteristic speeds for $g_2=0.005$. The evolution freezes inside black holes as a consequence of the $1+\log$ slicing condition that we use. Sounds horizons form for large initial data.}
\label{fig:weak_g2_pos_speeds}
\end{figure}

In Fig. \ref{fig:weak_g2_pos_det_h00} we display $\det(h^\m_{~\n})$ (top) and $-\a^2\,h^\m_{~\n}$ (bottom) for Cases 1--4. The white lines in these plots indicate the trajectories of the initial scalar field peak and serve to guide the eye.  In Cases 1 and 2, the scalar field bounces at the origin and eventually disperses to infinity; as the amplitude increases from Case 1 to Case 2, the scalar field spends more time near the origin where gravitational focusing is stronger. For sufficiently large amplitudes (Cases 3 and 4) it collapses into a black hole. In all cases, both $\det (h^\m_{~\n})>0$ and $-\a^2\, h^{00}>0$ throughout the evolution so the scalar equations are hyperbolic at all  times.  The long dashed line in Fig. \ref{fig:weak_g2_pos_det_h00} indicates the contour where the maximum of weak field condition \eqref{eq:WFC_2} is equal to one; as we can see, for Cases 1 and 2 the weak field condition is always less than one everywhere in spacetime, while in Cases 3 and 4 and the weak field condition is greater than one only inside the apparent horizon (solid black line). Only in Case 3 there is a small region near the origin where the weak field condition is greater than one and for a short period of time is not covered by an apparent horizon. Note however that this region is already cloaked by the sound horizon (dotted black line), so the scalar modes emanating from this region cannot reach asymptotic observers.   For Cases 2--4,    $\det(h^\m_{~\n})$ can significantly deviate from 1 (its GR value) when the scalar field is most contracted at the origin. Likewise, the bottom plots in Fig. \ref{fig:weak_g2_pos_det_h00} show that $-\a^2\, h^{00}$ also exhibits some deviation from its GR value near the origin but it never gets anywhere close to $0$.  Therefore, despite the weak field condition being small at all times, the Horndeski terms can have a significant impact on the dynamics of the system, especially near the origin where the gravitational focusing is strongest.

In Fig. \ref{fig:weak_g2_pos_speeds} we display the characteristic speeds for both the outgoing (top) and the ingoing (bottom) modes. Notice that in Case 2, both speeds approach zero at the origin when the scalar field collapses but their sign does not change.  This is indicative of strong gravitational dynamics, as one would expect since Case 2 is ``close" to the critical regime. Also, note that there are no scalar horizons in this case and all the scalar field eventually disperses to infinity. The dynamics changes in Cases 3 and 4, where a black hole forms. First, notice that $v_-$ changes sign inside the black hole, from negative to positive; this implies that inside the black hole, outgoing modes travel inwards, as expected. Eventually both speeds become close to zero in the region near the singularity. This is just a consequence of using $1+\log$ slicing in our simulations, which effectively freezes the evolution inside black holes. Second, we do observe the formation of scalar horizons, where $v_-=0$ and $v_+>0$.  
In both Cases 3 and 4, the characteristic speed of the outgoing modes is small in the vicinity of the sound horizon; consequently, even though the scalar field can eventually reach infinity, it will  remain near the black hole for a long time, thereby interacting with itself and with the black hole.  

It is apparent from the results shown here that even though the weak field condition \eqref{eq:WFC_2} is small everywhere, the scalar field still exhibits strong dynamics, such as the dynamical formation of scalar horizons. The  latter is a non-perturbative effect and it that can only be seen if one treats the Horndeski theory fully non-linearly. Evidently, if the couplings are small then the scalar horizon will be close to the metric horizon. In the case of a black hole binary in a Horndeski theory of gravity, even if the effects of the strong scalar dynamics are locally small, over a sufficiently long time they can lead to significant deviations from GR that may be observable \cite{PFTF2}. 

\subsubsection{Strong coupling}

\begin{figure}[t!]
\centering
\includegraphics[width=\textwidth]{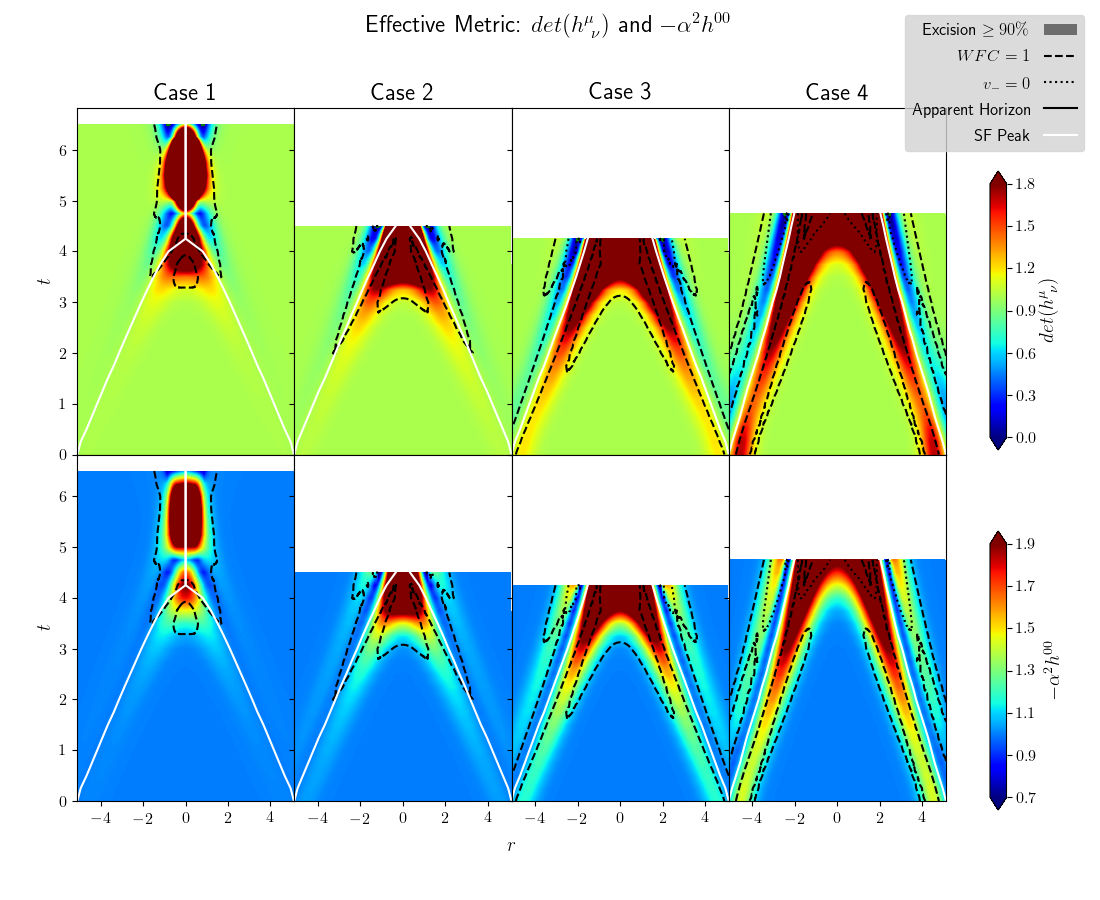}
\vspace{-7mm}
\caption{$\det (h^\m_{~\n})$ (\textit{top}) and $-\a^2\, h^{00}$ (\textit{bottom}) for strong and positive coupling $g_2=1.5$. The corresponding values of $\eta_2$, from left to right, are: $1.5\times 10^{-4},\,6\times 10^{-4},\,2.9\times 10^{-3},\,6.5\times 10^{-3}$.}
\label{fig:strong_g2_pos_det_h00}
\end{figure}

In this subsection we analyse the case for which the $g_2$ coupling is large and positive. We choose $g_2=1.5$ as a representative example. For this value of the coupling constant, the weak field condition \eqref{eq:WFC_2} can be $\mathcal{O}(1)$ for large initial data, see Fig. \ref{fig:strong_g2_pos_det_h00} Cases 3 and  4. Therefore, strictly speaking, in these cases the theory is already outside the regime of validity of EFT even though the initial value problem is well-posed. Nevertheless, we choose this value of the coupling constant as an illustrative example of the dynamics of Horndeski theories for large and positive $g_2$. 

%

In Fig. \ref{fig:strong_g2_pos_det_h00} we display  $\det(h^\m_{~\n})$ and $-\a^2 h^{00}$ during the evolution for our four cases. Unsurprisingly, this figure shows that in all cases the evolution breaks down at some point. For this choice of $g_2$ (and all other values of $g_2>0$), the reason why the simulations crash is because $\det(h^\m_{~\n})\to 0$ in a certain region at some instant of time and hence the scalar equation changes character, becoming parabolic. Beyond this point it is not possible to solve the equations as an initial value problem. For this value of the Horndeski coupling, for all Cases 1--4 the weak field condition \eqref{eq:WFC_2} has become large before the equations change character. Also, note that for large initial data (Cases 3 and 4), the evolution breaks down before an apparent horizon has had time to form and hence the pathology in the scalar equations of motion cannot be hidden behind the horizon.  Fig. \ref{fig:strong_g2_pos_det_h00} (bottom) shows that in all cases $-\a^2 h^{00}$ deviates significantly from its GR value and but remains well above zero up until the breakdown of the evolution. Likewise, we observe that in these simulations the characteristic speeds of both the ingoing and outgoing modes remain bounded at all times. Therefore, the loss of hyperbolicity for the  $g_2>0$ theories is due to a Tricomi-type-of transition, in accordance with the discussion in Section \ref{effective_metric_G2}.

By lowering the coupling constant a bit, it is possible to hide the strong scalar field dynamics that causes the breakdown of the hyperbolicity of the equations inside a large enough black hole. This is illustrated in Appendix \ref{appendix:other}, Fig. \ref{fig:intermdediate_g2_pos}. For such ``intermediate" couplings, the evolution still breaks down in Cases 2 and 3, while in Case 4 the pathologies that develop in the scalar equation can be hidden behind the horizon. In this case, one can continue the evolution without encountering any issues. Moreover, the weak field condition in Case 4 remains small on and  outside the black hole horizon despite the fact that $g_2$ is large. Clearly,  from the expression for the dimensionless coupling $\eta_2$, eq. \eqref{eq:eta2}, one can achieve the same results by increasing the initial amplitude $A$ instead of decreasing $g_2$. 

\subsubsection{Negative coupling}

\begin{figure}[t!]
\centering
\includegraphics[width=\textwidth]{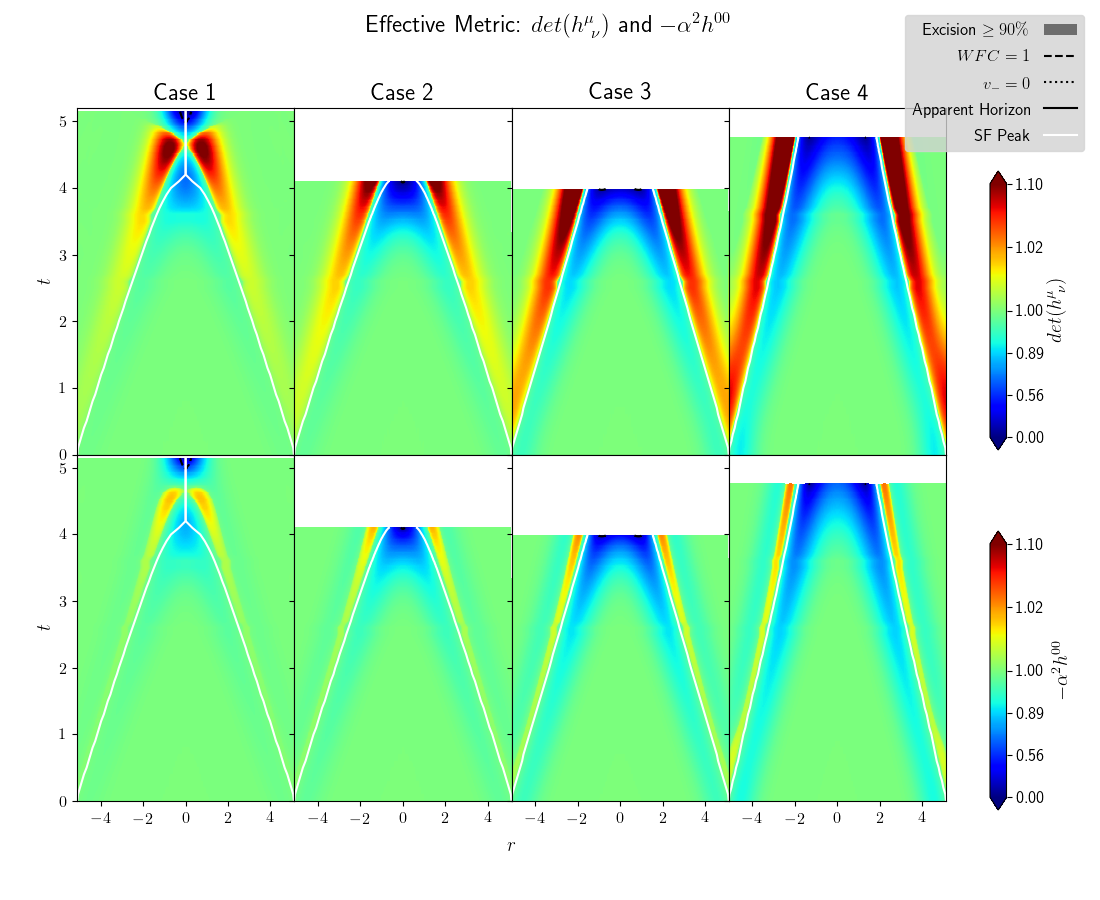}
\vspace{-7mm}
\caption{$\det (h^\m_{~\n})$ (\textit{top}) and $-\a^2\, h^{00}$ (\textit{bottom}) for a strong and negative coupling $g_2=-0.2$. The corresponding values of $\eta_2$, from left to right, are: $-2\times 10^{-5},-8\times 10^{-5},-3.9\times 10^{-4},-8.7\times 10^{-4}$.}
\label{fig:strong_g2_neg_det_h00}
\end{figure}

In this subsection we discuss the case of a strong and negative coupling constant $g_2$. As an illustrative example, we consider $g_2=-0.2$. 

\begin{figure}[t!]
\centering
\includegraphics[width=\textwidth]{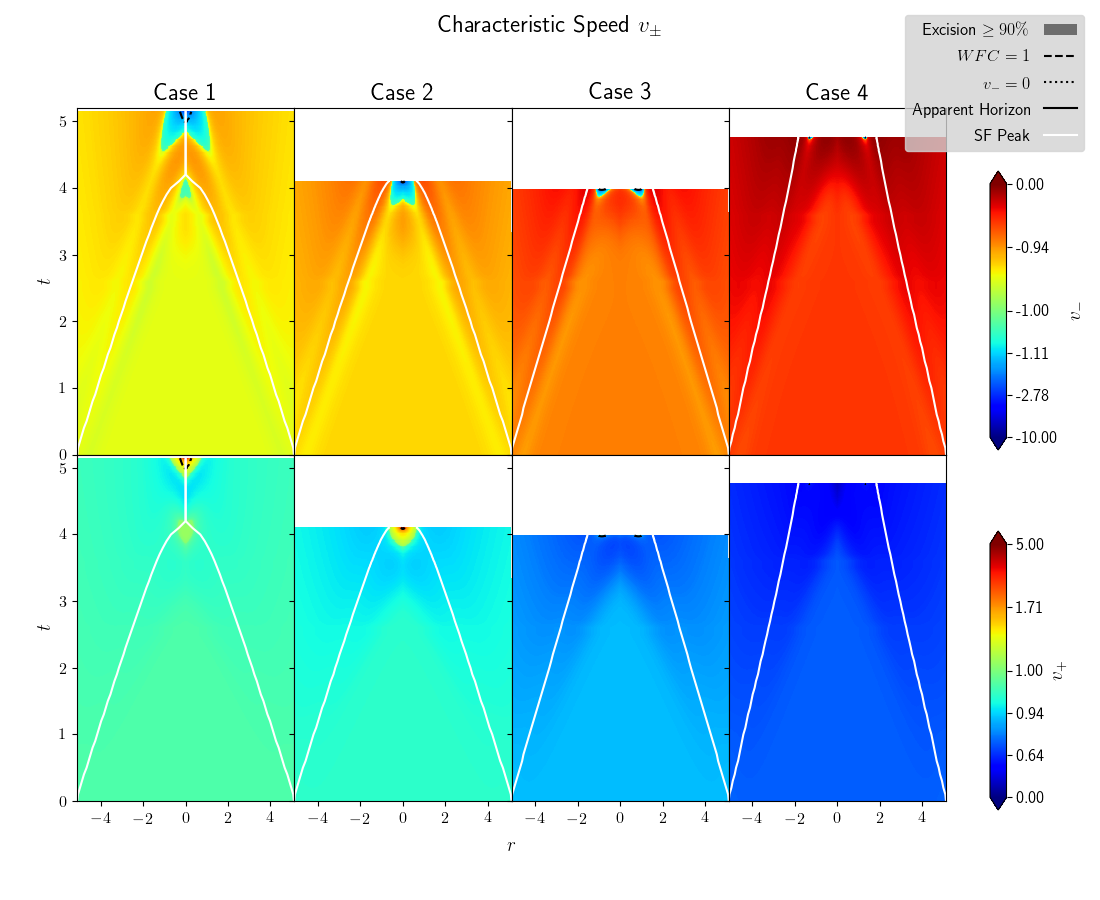}
\vspace{-7mm}
\caption{Characteristic speeds of the outgoing (\textit{top}) and ingoing (\textit{bottom}) scalar modes for $g_2=-0.2$. Both characteristic speeds simulatenously diverge when the evolution breaks down, but $v_-$ does so faster than $v_+$. This behaviour is in accordance with a Keldysh-type-of transition.}
\label{fig:fig:strong_g2_neg_speeds}
\end{figure}

As anticipated in Section \ref{effective_metric_G2}, the dynamics of the scalar field changes quite significantly for negative couplings. First, a smaller absolute value of $g_2$ is enough to cause a breakdown of the hyperbolicity of the scalar equations for both small and large data. The results are shown in Figs. \ref{fig:strong_g2_neg_det_h00} and \ref{fig:fig:strong_g2_neg_speeds}. In all cases we find that $-\a^2 h^{00}\to 0$ before $\det (h^\m_{~\n})\to 0$, even though this is not easily seen from Fig. \ref{fig:strong_g2_neg_det_h00}. This implies that, in our gauge, the $t=\textrm{const.}$ surfaces are no longer spacelike with respect to the scalar effective metric before the scalar equation changes character. The fact that for $g_2<0$, $-\a^2 h^{00}\to 0$ first results in infinite characteristic speeds of propagation for both the ingoing and outgoing modes, see Fig. \ref{fig:fig:strong_g2_neg_speeds}. Therefore, we associate the breakdown of the hyperbolicity of the scalar equation to a Keldysh-type-of transition, in accordance to the discussion in Section \ref{effective_metric_G2} (see also \cite{38_horndeski_new_may_example_luis}). The diverging characteristic speeds near the transition point imply that the dynamics of the scalar field becomes increasingly fast right before it breaks down; to adequately resolve it, in our simulations we had to significantly reduce the Courant factor. However, at some point  it is no longer feasible in practice to keep reducing it, and numerical errors eventually build up until the simulation inevitably crashes. A possible way out would be to change our slicing conditions to ensure that the $t=\textrm{const}.$ hypersurfaces remain spacelike with respect to both $h^{\mu\nu}$ and $g^{\mu\nu}$, but we have not attempted to do so here.

Notice that for this value of the coupling constant, the weak field condition \eqref{eq:WFC_2} is always less than one everywhere in spacetime, including the region near the origin where gravitational focusing is strongest, except immediately before the breakdown. This is simply a consequence of the fact that  $|\eta_2|$ is small in all Cases 1--4. Related to this last observation, the breakdown occurs before either sound horizons or apparent horizons have had time to form, so the pathologies cannot be hidden from asymptotic observers. However, when the breakdown occurs, even though the weak field condition \eqref{eq:WFC_2} may be as small as $\mathcal{O}(10^{-2})$, this is still much larger than $\eta_2$, thereby suggesting that the system is strongly coupled. We expect that a refined weak field condition should be able to capture that this case indeed becomes strongly coupled in a precise sense before the breakdown of the evolution. 

Needless to say, for sufficiently small absolute values of $|g_2|$ the  scalar equations remain hyperbolic at all times for Cases 1--4. In this situations the evolution is qualitatively similar to the small and positive $g_2$ case that we have already discussed in Subsection \ref{sec:g2_pos_weak_coupling}. Likewise, for a given $g_2<0$ and a sufficiently large $A$, the pathologies in the scalar equation can be hidden inside the black hole horizon.

\subsection{$G_3$ theories}
\label{sec:G3_neq_0}

\begin{figure}[t!]
    \centering
    \hspace*{-10mm}\includegraphics[width=1.15\textwidth]{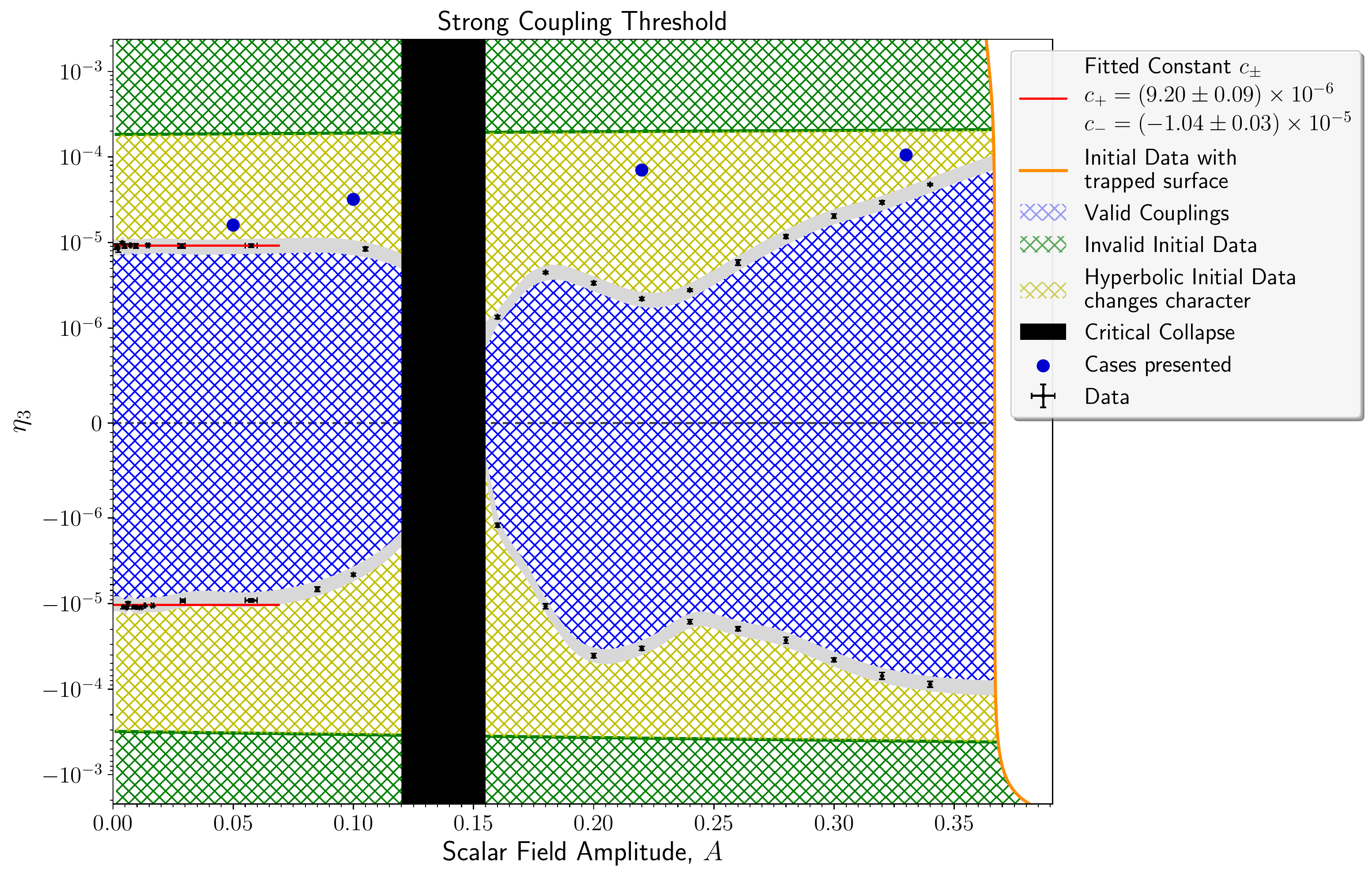}
    \caption{Dynamical regimes for the $G_3 = g_3\,X$ theory as a function of the initial amplitude $A$ and the dimensionless coupling constant $\eta_3$, see eq. \eqref{eq:eta3}. The black band denotes the region near critical collapse; black holes form to the right of this band. The orange curve on the right marks the region where the initial data contains a trapped surface. The scalar equation is hyperbolic at all times in the blue region; EFT is valid in the interior of this region. In the yellow region, the scalar equation is initially hyperbolic but it changes character during the evolution. In the green region the initial value problem is not well-posed.}
    \label{fig:results_eta3}
\end{figure}

In this subsection we will briefly comment the dynamics in Horndeski theories with $G_3 = g_3\,X$ and $G_2=0$. In all cases that we have explored, either for $g_3>0$ or $g_3<0$, the dynamics is qualitatively similar to the $G_2=g_2\, X^2$ theories with $g_2<0$ so we will not go into much detail. 

As discussed in Section \ref{effective_metric_G3}, we expect that for sufficiently small absolute values of $g_3$, the breakdown of the scalar evolution equations would be due to a Keldysh-type-of transition. Our numerical simulations confirm that this is indeed the case for either signs of $g_3$. In Figs. \ref{fig:strong_g3_pos_det_h00} and \ref{fig:strong_g3_pos_speeds} of Appendix \ref{appendix:other} we show the results for a representative case with $g_3=0.4$. In Fig. \ref{fig:strong_g3_pos_det_h00} we see that $-\a^2 h^{00}\to 0$ before $\det(h^\m_{~\nu})$ does, resulting in infinite characteristic speeds, as expected in a Keldsyh-type-of transition. In this case we observe that $v_-$ diverges as $-\a^2 h^{00}\to 0$ while  $v_+$ remains finite, see Fig. \ref{fig:strong_g3_pos_speeds}. Note that in this particular example, for large data (Cases 3 and 4) the evolution breaks down before the first apparent horizon appears. However, just as in the $G_2\neq 0$ theories, either by increasing the initial scalar amplitude so that a sufficiently large black hole forms or by lowering $|g_3|$, it is possible to hide the pathologies that may arise in the scalar evolution inside a black hole so that the theory remains in the regime of validity of EFT on and outside black holes. This is precisely what happens in the interior of the blue region in Fig. \ref{fig:results_eta3} in the large data regime.

Fig. \ref{fig:results_eta3} summarises our results for the $G_3=g_3\,X$ theories. The colour code is the same as in Fig. \ref{fig:results_eta2} and the qualitative features are also the same. The black band corresponds to the range of $A$ for which the future development of initial data becomes close to Choptuik's critical solution. Black holes form for $A$ to the right of the black band while for $A$'s to left, the scalar field disperses.  As before, global solutions to this particular Horndeski theory can be constructed for values of $(A,\eta_3)$ in the blue region. The regime of validity of EFT corresponds to the interior of the blue region, away from its boundaries. For $0<A\lesssim 0.05$, the boundary between the blue and yellow regions is at a constant value of $\eta_3$  given by $\eta_3\sim(9.20\pm 0.09)\times 10^{-6}$ and $\eta_3\sim(-1.04\pm 0.03)\times 10^{-5}$ respectively.

\section{Final remarks}
\label{conclusions}

In this paper we have studied the regime of validity of certain cubic Horndeski theories of gravity that have a well-posed initial value problem. We have chosen two particularly simple cases, namely \eqref{eq:relevantGs}, but we expect that our results should extend to other models as well, at least in the weakly coupled regime which is where these theories should be valid EFTs. For instance, for a single massive scalar field the results are qualitatively unchanged during gravitational collapse. Nevertheless, one expects that a massive scalar field will stay trapped around the black hole for a much longer time, forming scalar clouds \cite{Clough:2018exo}. Over long periods of time, such as in a black hole binary inspiral, the locally small deviations from GR introduced by Horndeski theories can (and will!) accumulate, giving rise to significant deviations. 

For the particular class of models that we have studied, the reason why the evolution breaks down is because the scalar equation changes character. For the $G_2=g_2\,X^2$ theory the transition can be of the Tricomi type for $g_2>0$, while for $g_2<0$ the transition is of the Keldysh type. On the other hand, for the $G_3=g_3\,X$ theory, we have only observed a breakdown \`a la Keldysh. However, this is not generic for the $G_3$ theories; other choices such as $G_3=g_3\,X^2$ can exhibit both behaviours. Furthermore, we have provided some level of analytic justification for the types of pathologies that may arise in each of the models that we have considered.  

In order for the initial value problem be well-posed and the theory be a consistent (truncated) EFT, we need to impose that a certain weak field condition \eqref{eq:WFC_2} is suitably small. For certain choices of initial conditions and couplings (no fine-tunning required) the conditions in \eqref{eq:WFC_2} can be $\mathcal{O}(1)$ and yet the scalar equation of motion is perfectly hyperbolic. Conversely, the conditions in \eqref{eq:WFC_2} can be $\mathcal{O}(10^{-2})$ and yet the scalar equation changes character. In either case, the weak field conditions at the time of breakdown are much larger than the dimensionless couplings, \eqref{eq:eta2}-\eqref{eq:eta3}, of the initial data. Therefore, in a certain sense, the theory becomes strongly coupled by the time the hyperbolicity is lost. It would be very interesting to obtain a sharp condition that identifies the truly weakly coupled regime of the theory and provides  some analytic understanding of it, at least for certain classes of initial data.

Having identified the regime regime of validity of the Hordneski theories that we have considered,  we can proceed to study black hole binaries for initial data in this regime. These studies will be presented in the companion paper \cite{PFTF2}.

\medskip

\section*{Acknowledgements}
We would like to thank Timothy Clifton, Aron D. Kovacs, Luis Lehner and Harvey S. Reall for discussions. We would also like to thank Harvey S. Reall and Eugeny Babichev for their insightful comments on an earlier version of the paper. Our special thanks are for the entire GRChombo collaboration (\texttt{www.grchombo.org}) for their help and support.  P.F. is supported by the European Research Council Grant No. ERC-2014-StG 639022-NewNGR, and by a Royal Society University Research Fellowship (Grant No. UF140319). P.F. and T.F. are supported by a Royal Society Enhancement Award (Grant No. RGF\textbackslash EA\textbackslash 180260). The simulations presented here were done on the MareNostrum4 cluster at the Barcelona Supercomputing Centre (Grant No. FI-2020-2-0011 and FI-2020-2-0016) and on the  Cambridge Service for Data Driven Discovery (CSD3), part of which is operated by the University of Cambridge Research Computing on behalf of the STFC DiRAC HPC Facility (www.dirac.ac.uk). The DiRAC component of CSD3 was funded by BEIS capital funding via STFC capital grants ST/P002307/1 and ST/R002452/1 and STFC operations grant ST/R00689X/1. DiRAC is part of the National e-Infrastructure.

P.F. would like to dedicate this work to his uncle Ramon Barnera, who spent 51 days in an  ICU with Covid-19 while this work was being completed. P.F. would like to express his gratitude to the public health workers and the tax payers who saved his uncle's life.

\newpage
\appendix 

\section{3+1 Conformal Decomposition}
\label{appendix:ccz4}

\subsection{Equations of Motion}
\label{appendix:EOM}

To carry out the numerical simulations presented in this paper, we used a code based on \texttt{GRChombo}, a multipurpose numerical relativity code \cite{grchombo}\footnote{See also \href{www.grchombo.org}{\texttt{www.grchombo.org}}.} that implements  the BBSNOK \cite{nakamura,nakamura2,shapiro2} or CCZ4 \cite{z4,z42,z4ori,zfourc,Z4bona} formulations of the Einstein equations. In this appendix we present the conformal 3+1 form of the stress tensor and the scalar equation \eqref{eq:scalar} as we have implemented in our code. 

Consider the usual timelike vector $n^\m$ normal to the spatial hypersurfaces; the projector $\gamma_{\m\n}=g_{\m\n}+ n_\m n_\n$ defines the spatial 3-metric $\gamma_{ij}$ with the corresponding covariant derivative $D_i$. From these, we obtain the following decomposition for the first derivatives of the scalar field:
\begin{align}
    \P &:= \mathcal{L}_n \f = n^\m \grad_\m \f\,, \label{eq:dtphi}\\
    \quad\quad\quad \P_i &:= D_i \f\,,
\end{align}
where $\mathcal{L}_n$ denotes the Lie derivative along $n^\m$. It follows that $\grad_\m\f=\P_\m - n_\m\P$ and $X=\tfrac{1}{2}\br{\P^2 - \P_i\P^i}$. We also decompose the second derivatives of the scalar field, defining the auxiliary variables:
\begin{gather}
\begin{aligned}
    \overline{\mathcal{L}_{\ve{n}}\Pi} &:= n^\mu n^\nu \grad_\mu \grad_\nu \phi = \mathcal{L}_{\ve{n}}\Pi - \Pi_{i} D^{i}\ln{\alpha}\,,\\
    \tau_{i} &:= \gamma_i^\mu n^\nu \grad_\mu \grad_\nu \phi = K_{ij} \Pi^{j} + D_{i}\Pi\,,\\
    \tau_{ij} &:= \gamma_i^\mu \gamma_j^\nu \grad_\mu \grad_\nu \phi = K_{ij} \Pi + D_{(i}\Pi_{j)}\,, 
    \label{eq:auxvars}
\end{aligned}
\end{gather}
and hence $\tau := \tau^{~i}_i = K\Pi + D^i\Pi_i$. Therefore, we get
\begin{align}
    \nabla_{\mu}\nabla_{\nu}\phi =&~ \overline{\mathcal{L}_{\ve{n}}\Pi}\,n_\mu\, n_\nu - 2\,n_{(\mu}\tau_{\nu)} + \tau_{\mu\nu}\,,\\
    \square\phi =&~ \tau - \overline{\mathcal{L}_{\ve{n}}\Pi}\,,
\end{align}
with $n^\mu \tau_\mu = 0$ and $n^\mu \tau_{\mu\nu}=0$.

In terms of the usual conformal spatial metric $\ti{\gamma}_{ij}:=\chi\gamma_{ij}$ (with $\det(\ti{\gamma}_{ij})=1$) and its associated covariant derivative $\ti{D}_i$, we define the conformal variables for the scalar field as,
\begin{gather}
\begin{aligned}
    \ti{\P}_i &:= \ti{D}_i\f\,, \quad\quad\quad \ti{\tau}_{i} &:= \tau_i\,, \quad\quad\quad \ti{\tau}_{ij} &:= \chi\tau_{ij}\,.
\end{aligned}
\end{gather}
Note that the indices of $\ti{\tau}_{ij}$ are raised with the conformal metric $\ti{\gamma}_{ij}$ so that $\ti{\t} := \ti{\t}^{~i}_i = \t$, and similarly for all other conformal variables. For example, $\ti{\P}^i = \frac{1}{\chi}\P^i$, which implies $X = \tfrac{1}{2}(\Pi^2 - \chi\ti{\Pi}_i\ti{\Pi}^i)$. With these definitions in place, the 3+1 conformal decomposition of the scalar energy-momentum tensor is:
\begin{align}
    \kappa \rho :=&~ \kappa\,n^\m n^\n T_{\m\n} \nonumber \\
    =&~ V - G_2 + \tfrac{1}{2}\big(\Pi^2 + \chi\ti{\Pi}_{i} \ti{\Pi}^{i}\big)\br{1+2\partial_\phi G_3} + \partial_X G_3(\ti{\tau} \Pi^2 -  \chi\ti{\Pi}^{i} \ti{\Pi}^{j} \ti{\tau}_{ij})  + \Pi^2 \partial_X G_2\,, \\
    \kappa S_{i} :=& -\kappa\,n^\m \g_i^{~n} T_{\m\n} \nonumber \\
    =& -\Pi~\ti{\Pi}_{i} \bigl(1 + \partial_X G_2 + 2 \partial_\phi G_3\bigr) + \partial_X G_3(\chi\ti{\Pi}_{i} \ti{\Pi}^{j} \ti{\tau}_{j} + \Pi ~\ti{\Pi}^{j} \ti{\tau}_{ij} - \ti{\tau} \Pi ~\ti{\Pi}_{i} -  \Pi^2 \ti{\tau}_{i})\,,\\
    \kappa S_{ij} :=&~ \kappa\,\g_i^{~\m} \g_j^{\n}T_{\m\n} \nonumber \\
    =&~ \ti{\Pi}_{i} \ti{\Pi}_{j} \bigl(1 + \partial_X G_2 + 2 \partial_\phi G_3\bigr) + \tfrac{1}{\chi}\ti{\gamma}_{ij} \bigl(G_2 - V + X + 2 X \partial_\phi G_3\bigr)\\
    & +\partial_X G_3 \Big[\ti{\tau} \ti{\Pi}_{i} \ti{\Pi}_{j} + 2\Pi~ \ti{\Pi}_{(i} \ti{\tau}_{j)} - 2\ti{\Pi}^{k}\ti{\Pi}_{(i} \ti{\tau}_{j)k} -  \ti{\gamma}_{ij} \ti{\Pi}^{k} (2 \Pi \ti{\tau}_{k} -  \ti{\Pi}^{l} \ti{\tau}_{kl}) \nonumber \\
    &\hspace{1.8cm}+ \overline{\mathcal{L}_{\ve{n}}\Pi}(\tfrac{1}{\chi}\ti{\gamma}_{ij}\Pi^2 - \ti{\Pi}_i\ti{\Pi}_j)\Big]\,, \nonumber
\end{align}
Similarly, the scalar field evolution equation \eqref{eq:scalar} in first order form is given by \eqref{eq:dtphi} and:
\begin{equation}
\begin{aligned}
    &\overline{\mathcal{L}_{\ve{n}}\Pi}\Big[1 + \partial_X G_2 + 2 \partial_\phi G_3 + 2 \ti{\tau} \partial_X G_3 - X^2 \bigl(\partial_X G_3\bigr)^2 - \chi\ti{\Pi}^{i} \ti{\Pi}^{j} \ti{\tau}_{ij} \partial^2_{XX} G_3 - 2X\partial^2_{\phi X} G_3 \\
    &~~~~~~~ + \Pi^2\br{2X\bigl(\partial_X G_3\bigr)^2 + \partial^2_{XX} G_2 + \ti{\tau}\partial^2_{XX} G_3 + 2\partial^2_{\phi X} G_3}\Big] =\\
    =~& \partial_\phi G_2 - \partial_\phi V + \ti{\tau} \sbr{1 + \partial_X G_2 + 2 \partial_\phi G_3 + \ti{\tau} \partial_X G_3 - X^2\bigl(\partial_X G_3\bigr)^2 -2X\partial^2_{\phi X} G_3} \\
    &+\sbr{\partial^2_{XX} G_2 + 2\partial^2_{\phi X} G_3 + 2X\bigl(\partial_X G_3\bigr)^2 + \ti{\tau} \partial^2_{XX} G_3}\chi(2\Pi~ \ti{\Pi}^{i} \ti{\tau}_{i} - \ti{\Pi}^{i} \ti{\Pi}^{j} \ti{\tau}_{ij}) \\
    &- (\partial^2_{\phi X} G_2 + \partial^2_{\phi\phi}G_3) 2 X \\
    &+ \chi\partial^2_{XX} G_3\sbr{(\Pi\ti{\tau}_i - \ti{\Pi}^j\ti{\tau}_{ji})(\Pi\ti{\tau}^i - \ti{\Pi}_k\ti{\tau}^{ki}) - \chi \ti{\Pi}^{i} \ti{\Pi}^{j} \ti{\tau}_{i} \ti{\tau}_{j}} \\\label{eq:Horndeski_final}
    &-\partial_X G_3 \sbr{ G_2 X - 2\chi \ti{\tau}_{i} \ti{\tau}^{i} + \ti{\tau}_{ij} \ti{\tau}^{ij} + X^2 (2 + \partial_X G_2 + 4 \partial_\phi G_3)}\,.
\end{aligned}
\end{equation}

Note that one can obtain the standard 3+1 evolution equations without a conformal transformation by setting $\chi=1$ and dropping any ` $\ti{}$ ' superscripts.

Regarding gauge and numerical evolution parameters, we choose $1+\log$ slicing and hyperbolic gamma-driver condition with the standard parameters. We use CCZ4 paramaters $\{\kappa_1 = \tfrac{0.1}{\alpha}$, $\kappa_2 = 0$, $\kappa_3 = 1\}$ and Kreiss-Oliger numerical dissipation with $\sigma=0.3$. Typical simulations used a Courant factor of $0.2$ (reduced for Keldysh-type-of transitions), a coarse grid resolution of $\D x = 1$ and up to 7 additional refinement levels, and a box size of $L = 96$ with Sommerfeld boundary conditions. We use the gradients $\f$ and $\chi$ as well as contours of $\chi$ to tag cells for regridding. Last but not least, we use the symmetry of the system to only simulate one octant of the full domain, which reduces the computational cost of the problem.

\subsection{Effective metric}

As discussed in Section \ref{sec:effmetric}, the quantities $-\a^2 h^{00}$ and $\text{det}\br{h^\m_{~\n}}$ are useful to monitor the hyperbolicity of the scalar equation of motion and determine whether its change of character is of the  Tricomi or Keldysh type. Here we present $-\a^2 h^{00}$ and $h^\m_{~\n}$ in terms of the 3+1 conformal variables, which is how we have calculated them in our code:
\begin{equation}
\begin{aligned}
    h^0_{~i} =&~ \frac{1}{\alpha}\Big\{\Pi~\ti{\Pi}_i\Big[2\,X\,\bigl(\partial_X G_3\bigr)^2 + \partial^2_{XX} G_2 + \ti{\tau}\,\partial^2_{XX} G_3 + 2\,\partial^2_{\phi X} G_3\Big] \\
        &\hspace{0.7cm} + \ti{\tau}_i\br{2\,\partial_X G_3 + \Pi^2\, \partial^2_{XX}G_3}-\partial^2_{XX}G_3\br{\ti{\Pi}^k\,\ti{\tau}_{ki}\,\Pi + \chi\,\ti{\Pi}_i\,\ti{\Pi}^k\,\ti{\tau}_k} \Big\}\,,\\
    h^0_{~0} =&~ \beta^k h^0_{~k} -\a^2 h^{00} \,,\\
    h^i_{~j} =&~ -\beta^i h^0_{~j} - \chi\,\ti{\Pi}^i\,\ti{\Pi}_j\Big[2X\bigl(\partial_X G_3\bigr)^2 + \partial^2_{XX} G_2 + \ti{\tau}\,\partial^2_{XX} G_3 + 2\,\partial^2_{\phi X} G_3\Big]\\
        & + \delta^i_{~j}\Big[ 1 + \partial_X G_2 + 2\, \partial_\phi G_3 + 2\,\ti{\tau}\, \partial_X G_3 - X^2 \bigl(\partial_X G_3\bigr)^2 
         - 2\,X\,\partial^2_{\phi X} G_3\\
        &\hspace{1.2cm} - \chi\,\ti{\Pi}^{k}\, \ti{\Pi}^{l}\, \ti{\tau}_{kl}\, \partial^2_{XX} G_3 + 2\,\chi\,\Pi~\ti{\Pi}^k\,\ti{\tau}_k\,\partial^2_{XX}G_3\Big] \\
        &- 2\,\partial_X G_3\,\ti{\tau}^i_{~j} + \chi\,\partial^2_{XX}G_3\br{\ti{\Pi}^{i}\,\ti{\tau}_{jk}\,\ti{\Pi}^k + \ti{\Pi}_{j}\,\ti{\tau}^{ik}\,\ti{\Pi}_k - \Pi~\ti{\Pi}^{i}\ti{\tau}_{j} - \Pi~\ti{\Pi}_{j}\,\ti{\tau}^{i}}\\
        & -\overline{\mathcal{L}_{\ve{n}}\Pi}\sbr{\delta^i_{~j}\br{2\,\partial_X G_3 + \Pi^2 \partial^2_{XX} G_3} - \chi\,\ti{\Pi}^i\,\ti{\Pi}_j\,\partial^2_{XX}G_3} \,, \\
    h^i_{~0} =&~ \beta^k h^i_{~k } - \alpha^2 \chi\ti{\gamma}^{ik}h^0_{~k} + \a^2 h^{00}\,\beta^i\,.
\end{aligned}
\label{eq:hud}
\end{equation}
and,
\begin{equation}
\begin{aligned}
    -\a^2 h^{00} =&~ 1 + \partial_X G_2 + 2\, \partial_\phi G_3 + 2\,\ti{\tau}\, \partial_X G_3 - X^2 \bigl(\partial_X G_3\bigr)^2 \\
    &- \chi\,\ti{\Pi}^{i}\, \ti{\Pi}^{j}\, \ti{\tau}_{ij}\, \partial^2_{XX} G_3 - 2X\partial^2_{\phi X} G_3 \\
        & + \Pi^2\big[2\,X\bigl(\partial_X G_3\bigr)^2 + \partial^2_{XX} G_2 + \ti{\tau}\,\partial^2_{XX} G_3 + 2\,\partial^2_{\phi X} G_3\big]\,,
\end{aligned}
\end{equation}

From \eqref{eq:hud} one can readily compute $\det(h^\m_{~\n})$. If necessary, the effective metric with both indices up can also be obtained by raising the lower index in \eqref{eq:hud} with the spacetime metric.

\section{Determinant of the effective metric}
\label{appendix:det_eff_metric}

We can compute $\text{det}\br{h^\m_{~\n}}$ in full generality using Cayley–Hamilton's theorem and Newton's identities. The general case, with both $G_2\neq 0$ and $G_3\neq 0$, is not particularly insightful and in practice it is preferable to directly compute the  determinant of the metric with a lowered index numerically. For clarity, in this Appendix we provide the explicit expression for the determinant in the case $G_2=0$ and $G_3=g_3\,X$:

\begin{gather}\nonumber
\begin{aligned}
    \text{det}\br{h^\m_{~\n}} =&~ 1 + 6g_3\square\f\,+\\
    & + g_3^2\big[14\br{\Box\f}^2-2\br{\grad_\m\grad_\n\f}\br{\grad^\m\grad^\n\f}\big]\\
    & + g_3^3\bigg[\frac{44}{3}\br{\Box\f}^3 - 2\Box\phi\br{2\br{\grad_\m\grad_\n\f}\br{\grad^\m\grad^\n\f} + X^2}\\
    &\hspace{1cm}-4X\br{\grad^\m\f}\br{\grad^\n\f}\br{\grad_\m\grad_\n\f}
      - \frac{8}{3}\br{\grad_\m\grad_\n\f}\br{\grad^\m\grad^\r\f}\br{\grad^\n\grad_\r\f}\bigg]
\end{aligned}\\
\begin{aligned}
    & + g_3^4\Big[6\br{\Box\f}^4-4\br{\Box\f}^2 \br{X^2+\br{\grad^\m\grad^\n\f}\br{\grad_\m\grad_\n\f}} - 6X^4 \\
    &\hspace{1cm}-8\Box\f X \br{\grad^\m\f}\br{\grad^\n\f}\br{\grad_\m\grad_\n\f} 
    -8X\br{\grad^\m\f}\br{\grad^\n\f}\br{\grad_\m\grad_\r\f}\br{\grad^\r\grad_\n\f}\\
    &\hspace{1cm} -4\br{\grad^\m\grad_\n\f}\br{\grad^\n\grad_\r\f}\br{\grad^\r\grad_\s\f}\br{\grad^\s\grad_\m\f}\\
    & \hspace{1cm} + \br{\grad^\m\grad^\n\f}\br{\grad_\m\grad_\n\f}\br{-4X^2+2\br{\grad^\r\grad^\s\f}\br{\grad_\r\grad_\s\f}}\Big]\\
    & + g_3^5\Big[2\Box\f X^2\br{2\br{\grad^\m\grad^\n\f}\br{\grad_\m\grad_\n\f}-7X^2}-4\br{\Box\f}^3 X^2 \\
    & \hspace{1cm} + 8X^2\br{\grad^\m\grad_\n\f}\br{\grad^\n\grad_\r\f}\br{\grad^\r\grad_\m\f}\\
    & \hspace{1cm}-16X\br{\grad^\m\f}\br{\grad^\n\f}\br{\grad^\r\grad_\m\f}\br{\grad^\s\grad_\n\f}\br{\grad_\r\grad_\s\f}\\
    & \hspace{1cm} + 8X \br{\grad^\m\f}\br{\grad^\n\f}\br{\grad_\m\grad_\n\f}\br{-\br{\Box\f}^2+ X^2 + \br{\grad^\r\grad^\s\f}\br{\grad_\r\grad_\s\f}}\Big]\\
    & + g_3^6\Big[8\Box\f X^3 \br{\grad^\m\f}\br{\grad^\n\f}\br{\grad_\m\grad_\n\f} \\
    &\hspace{1cm}+ 3X^4 \br{2\br{\grad^\m\grad^\n\f}\br{\grad_\m\grad_\n\f}-5\br{\Box\f}^2} \\
    & \hspace{1cm} + 8X^6 + 8X^3\br{\grad^\m\f}\br{\grad^\n\f}\br{\grad_\m\grad_\r\f}\br{\grad_\n\grad^\r\f} \Big]\\
    & + g_3^7 X^5\br{10\Box\f X - 4 \br{\grad^\m\f}\br{\grad^\n\f}\br{\grad_\m\grad_\n\f}} - 3g_3^8 X^8\,.
\end{aligned}
\label{eq:det_full_g3}
\end{gather}

\section{Dealing with strong field regime inside black holes}
\label{appendix:excision}

As described in  Section \ref{summary}, inevitably the evolution will exit the regime of validity of Horndeski theories inside black holes. To deal with this situation, in practice we excise a portion of the interior of the black hole. In this appendix we provide the details of our implementation. 

Rather than performing proper excision, i.e., cutting out a region of the domain, we found that it was easier to modify the evolution equations inside the black hole. The result should be the same as information cannot escape from this region. Note that in certain Horndeski theories, depending on the sign of the couplings, the scalar field can propagate faster than light and consequently the associated scalar apparent horizon will be inside the black hole horizon \cite{67_G2Horndeski_gtt_to_zero}. Therefore, to avoid unphysical effects leaking out of the black hole,  any modification of the equations of motion should be done in a region contained within all apparent horizons.

Since puncture gauge can handle singularities very well in GR, in practice we turn off all Horndeski terms in a certain region inside the black hole and evolve the standard GR equations there. To do so, we first define a smooth transition function, valued between 0 and 1, as the sigmoid-like function:
\begin{equation}
    \sigma(x; \bar{x}, w) = \frac{1}{1 + e^{-\frac{2}{w}\br{\frac{x}{\bar{x}}-1}}}\,,
\end{equation}
where $\bar{x}$ represents the transition point, and $w$ represents the transition width, relative to $\bar{x}$, such that $w\bar{x}$ is the actual width of the transition.\footnote{Roughly, $\sigma${ \footnotesize$\lesssim$ }$0.1$ for $x<\bar{x}(1-w)$ and $\sigma${ \footnotesize$\gtrsim$ }$0.9$ for $x>\bar{x}(1+w)$, and $\sigma$ decays very fast to 0 or 1 outside of this interval.} The metric apparent horizon can be accurately tracked during simulations, but for a sense of what is ``well within the black hole", contours of the conformal factor $\chi$ are, in puncture gauge, an excellent measure. For example, for a Schwarzschild black hole, after puncture gauge settles, the apparent horizon corresponds to a contour of $\chi$ around $0.25$, reducing to lower values as spin increases along the Kerr family of solutions. For reasonable choices of Horndeski couplings in the regime of validity of the theory, the scalar apparent horizon is close to the metric horizon. Therefore,  the region inside a certain sufficiently small contour of $\chi$ should be contained in all apparent horizons.  Denoting by $W$ the maximum of all the weak field conditions \eqref{eq:WFC_1},  we  define the excision function $e(\chi,W)$ as:
\begin{equation}
    e(\chi,W) = \sigma(\chi; \bar{\chi}, -w_\chi)\,\sigma(W; \bar{W}, w_W)\,,
    \label{eq:excision}
\end{equation}
where $w_\chi$ and $w_W$ are two adjustable parameters. In our simulations we typically used $\bar{\chi}=0.1$, $w_\chi = 0.2$, $\bar{W}=1$, $w_W=0.1$. This choice is robust, in the sense that changing these barely affects the evolution across resolutions as long as $\bar{\chi}$ is well within the black hole, which is the case for this choice. It follows from the definition \eqref{eq:excision} that $e\to 1$ when $\chi<\bar{\chi}$ and $W>\bar{W}$, and $e\to0$ otherwise.  We then modify the right hand side of the evolution equations, collectively denoted by RHS, as:
\begin{equation}
    \text{RHS} = (1-e)\text{RHS}_\text{Horndeski} + e~\text{RHS}_\text{GR}\,.
\end{equation}
with $e$ given by \eqref{eq:excision}. In practice, we are only modifying the equations of motion in a region where the weak field condition is large and where the theory should not be trusted anyway.

\section{Convergence}
\label{appendix:convergence}

%
\begin{figure}[ht!]
\centering
\includegraphics[width=1\textwidth]{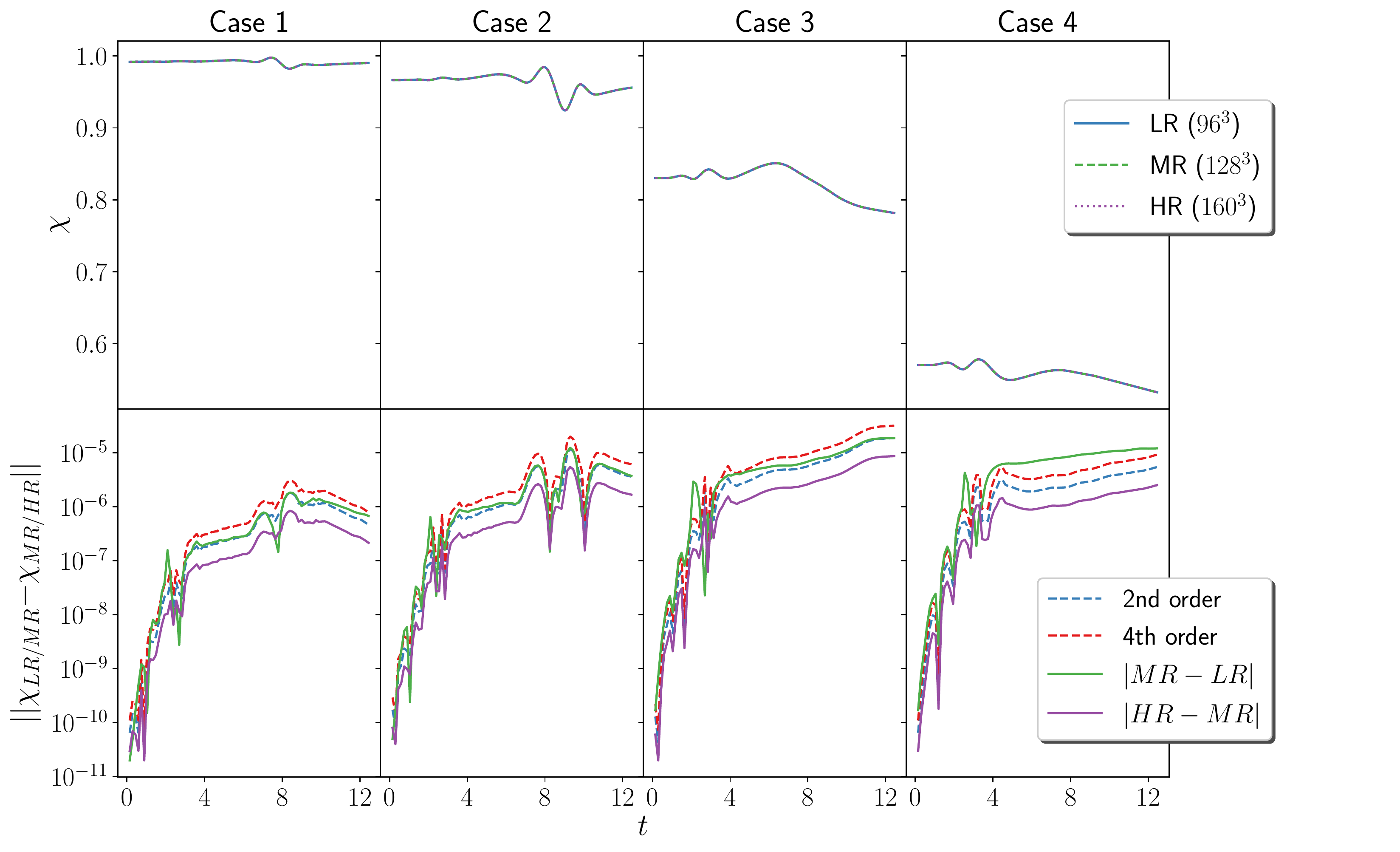}
\vspace{-5mm}
\caption{Convergence test for the $g_2=0.005$ run with different coarse resolutions: low ($LR$: $96^3$), medium ($MR$: $128^3$) and high ($HR$: $160^3$) resolutions, in addition to 7 refinement levels. \textit{Top}: evolution of $\chi$ at a fixed radius of $r=3$. \textit{Bottom}: $|MR-LR|$ and $|HR-MR|$ errors and the expected values for $|MR-LR|$ assuming $2^{\text{nd}}$ and $4^{\text{th}}$ order convergence.}
\label{fig:convergence}
\end{figure}
%
%

In this appendix we provide  details of some of the convergence tests that we have carried out. As an illustrative example, we consider the weak coupling $g_2=0.005$ case presented in \ref{sec:g2_pos_weak_coupling}. To carry out the convergence tests, we used simulations with coarsest level resolutions $\D x = 1$ (low resolution, $LR$), $\D x = 0.75$ (medium resolution, $MR$) and $\D x = 0.6$ (high resolution, $HR$) respectively, all with the same 7 additional levels of refinement. The results of the simulations for the 4 cases analysed are shown in Fig. \ref{fig:convergence}.  The bottom panel shows the error estimates $|MR-LR|$ (solid green curve) and  $|HR-MR|$ (solid purple curve), and compares them to the expected errors for  $2^{\text{nd}}$ (dashed blue) and $4^{\text{th}}$ (dashed red) order convergence. The latter were obtained from  the $|HR-MR|$ error using the continuum limit of the convergence factor: $\frac{\br{\D x_{LR}}^n - \br{\D x_{MR}}^n}{\br{\D x_{MR}}^n - \br{\D x_{HR}}^n}$. We see that our numerical results are consistent with convergence order between 2 and 4. Notice that it appears that the evolution has not reached a stationary state, but this should not be a concern since the outcome in terms of well-posedness and possible pathologies has already been determined after collapse occured.

We also monitor the behaviour of the Hamiltonian and Momentum constraints for the simulation with $g_2=0.005$ presented in \ref{sec:g2_pos_weak_coupling}. We measure the $L^2$ norm of a quantity $\mathcal{Q}$ by the volume average:
\begin{equation}
	L^2\mathcal{Q} = \sqrt{\frac{1}{V}\int_V|\mathcal{Q}^2|dV}\,,
\end{equation}
where $V$ is the volume of the box except the region excised inside black holes (if there are any present). 

\begin{figure}[t]
\centering
\includegraphics[width=0.99\textwidth]{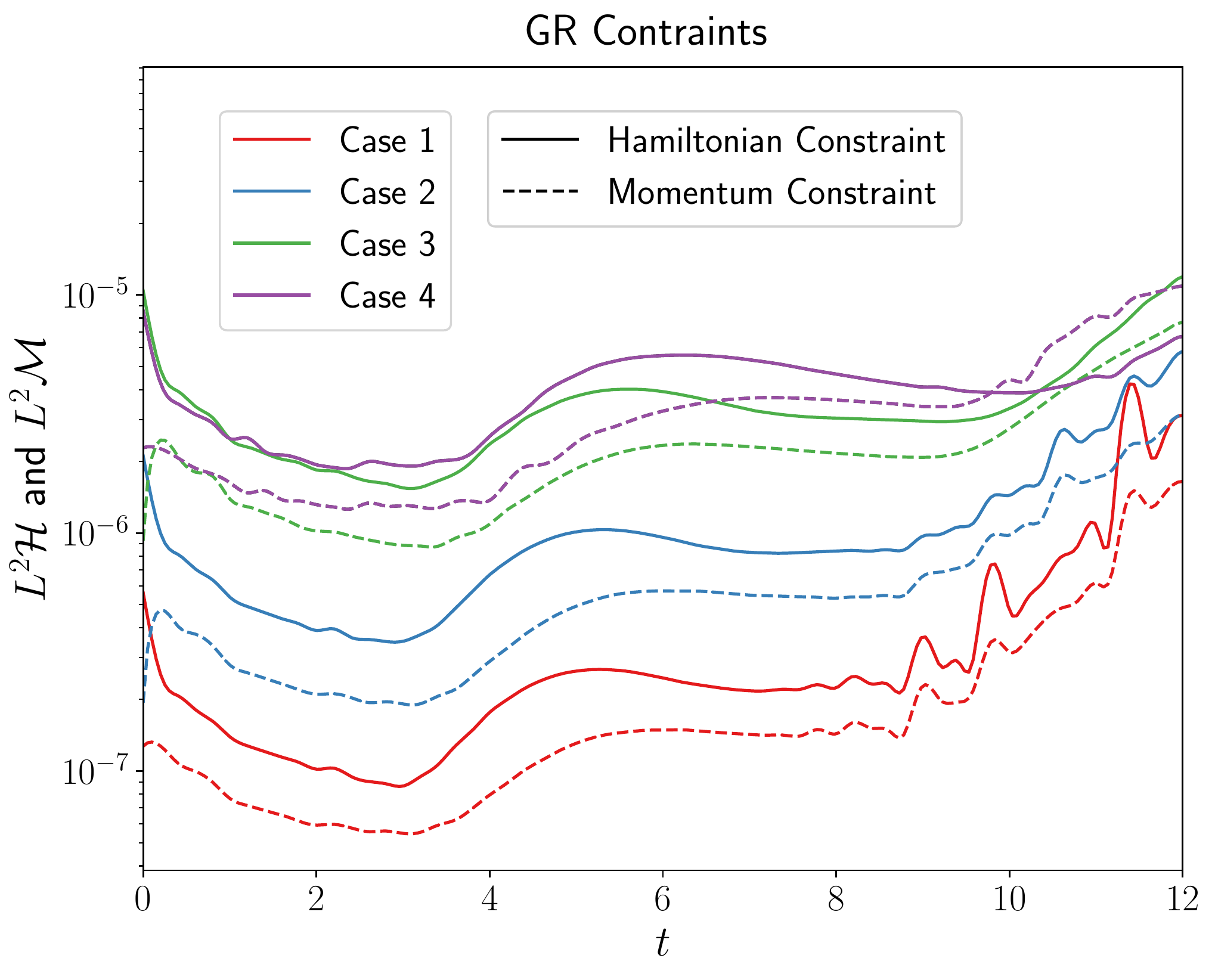}
\vspace{-2mm}
\caption{$L^2$ norm of constraints for the $g_2=0.005$ run  with coarsest level resolution of $\D x=1$ and 7 additional levels of refinement.}
\label{fig:constraints}
\end{figure}

In Fig. \ref{fig:constraints} we show the $L^2$ norms of the Hamiltonian constraint $\mathcal{H}$ and the Euclidean norm of the momentum constraints $\mathcal{M}\equiv ||\mathbf{M}||$. This plot gives some information about the absolute level of error in our simulation and it shows that is below $10^{-5}$ throughout the evolution. Note that the constraint violations seem to increase at late times. The reason is because  some of the scalar field (or all of it in Cases 1 and 2) disperses to infinity; as the scalar field propagates towards the outer boundaries, it moves away from the center of the grid into coarser refinement levels, and thus resolution is lost.

\begin{figure}[!ht]
\centering
\includegraphics[width=0.99\textwidth]{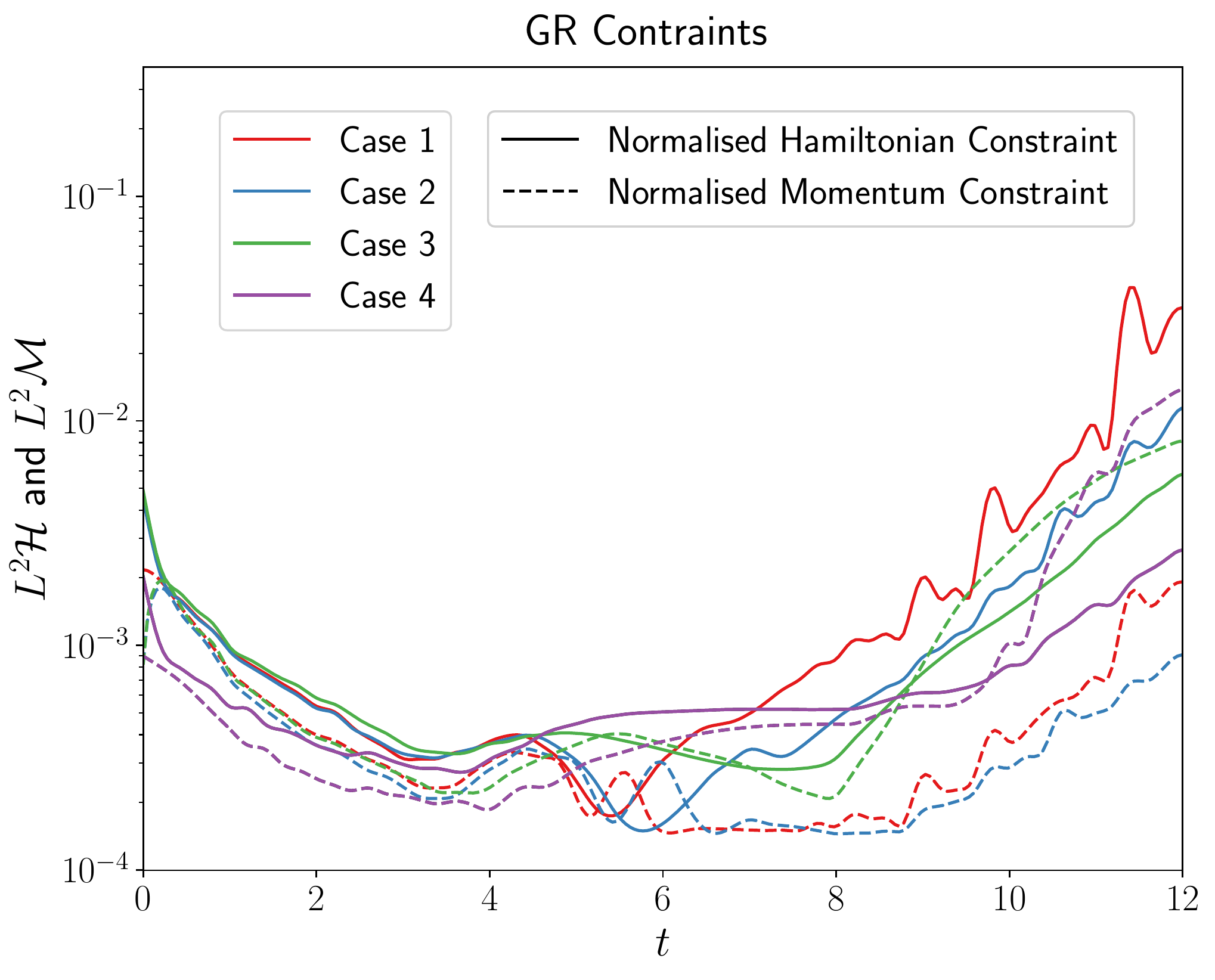}
\vspace{-2mm}
\caption{$L^2$ norm of the normalised constraints for the $g_2=0.005$ run  with coarsest level resolution of $\D x=1$ and 7 additional levels of refinement. During gravitational collapse the error is under the $0.1\%$ level; the increase of the relative errors in the later stages of the evolution is due to the fact that the normalisation factors themselves become very small.}
\label{fig:constraints_norm}
\end{figure}

To obtain a more useful insight about the relative errors, in Fig. \ref{fig:constraints_norm} we consider the normalised  $L^2$ norm of the constraints, which is dimensionless.  In more detail,  we normalise the $L^2$ norm of a given constraint $\mathcal{Q}$ by the $L^2$ norm of the sum of the absolute value of each term in the expression for $\mathcal{Q}$; for instance, in the case of the Hamiltonian constraint, 
\begin{equation}
\mathcal{H}=R + \frac{2}{3}K^2 - A_{ij}A^{ij} - \k\r\,,
\end{equation} 
the normalisation factor that we use is  the $L^2$ norm of $|R| + \left|\frac{2}{3}K^2\right| + \left|A_{ij}A^{ij}\right| + \left|\k\r\right|$. Fig. \ref{fig:constraints_norm} shows that constraint violations are under the $0.1\%$ level during gravitational collapse. At late times, as the scalar field disperses or is absorbed by the black hole, matter terms in the constraints become increasingly small and, as a consequence, the normalisation factors also significantly decrease; in turn, this also leads to an increase of the normalised constraint. Therefore, we can conclude that we have a good numerical control over our simulations. 

\clearpage

\section{Other cases of interest}
\label{appendix:other}

\begin{figure}[H]
\centering
\includegraphics[width=1\textwidth]{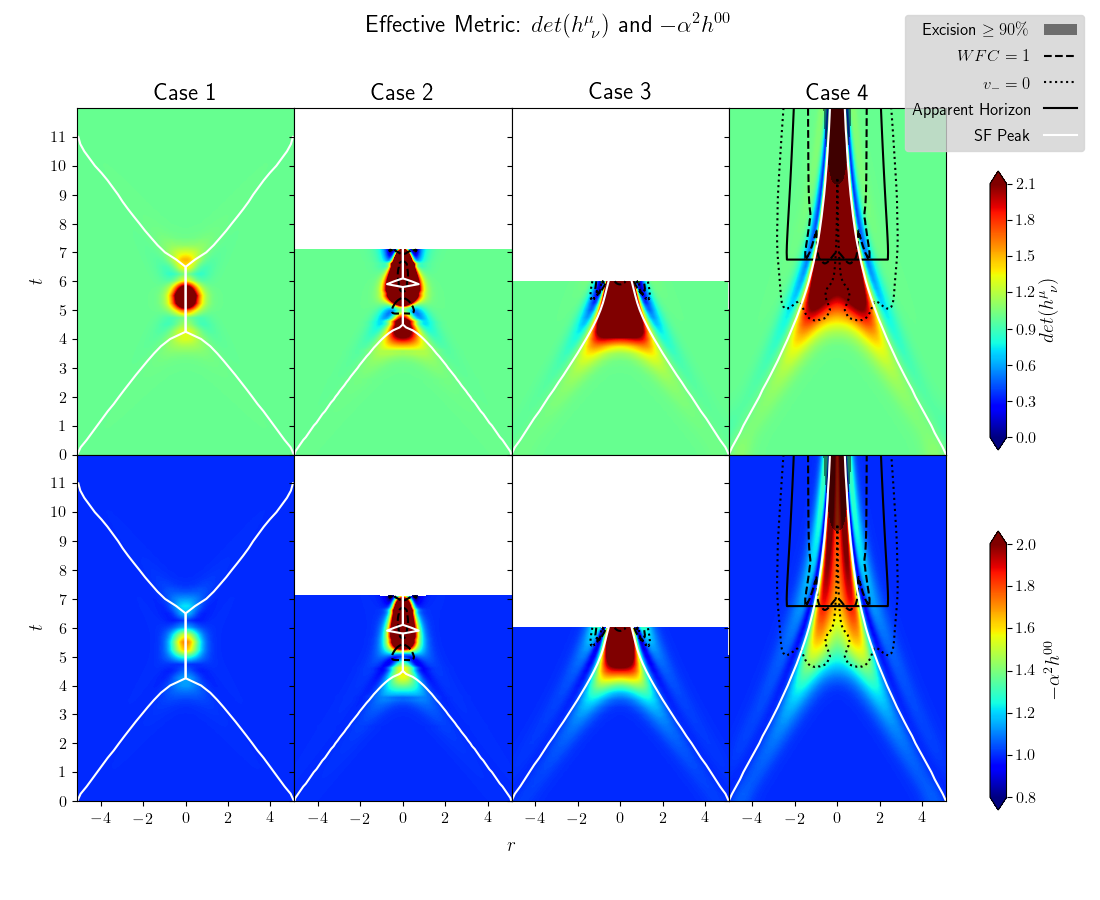}
\caption{$\det (h^\m_{~\n})$ (\textit{top}) and $-\a^2\, h^{00}$ (\textit{bottom}) for an intermediate positive coupling: $g_2=0.2$. The corresponding values of $\eta_2$, from left to right, are: $2\times 10^{-5},8\times 10^{-5},3.9\times 10^{-4},8.7\times 10^{-4}$. For small enough initial data (Case 1) the evolution is perfectly consistent, while it breaks down in a Tricomi-type of transition in Cases 2 and 3. For large enough initial data (Case 4), the pathologies that may develop during the evolution are hidden behind the black hole horizon. In this case,  the weak field condition is small on and outside the black hole horizon.}
\label{fig:intermdediate_g2_pos}
\end{figure}

\begin{figure}[H]
\centering
\includegraphics[width=1\textwidth]{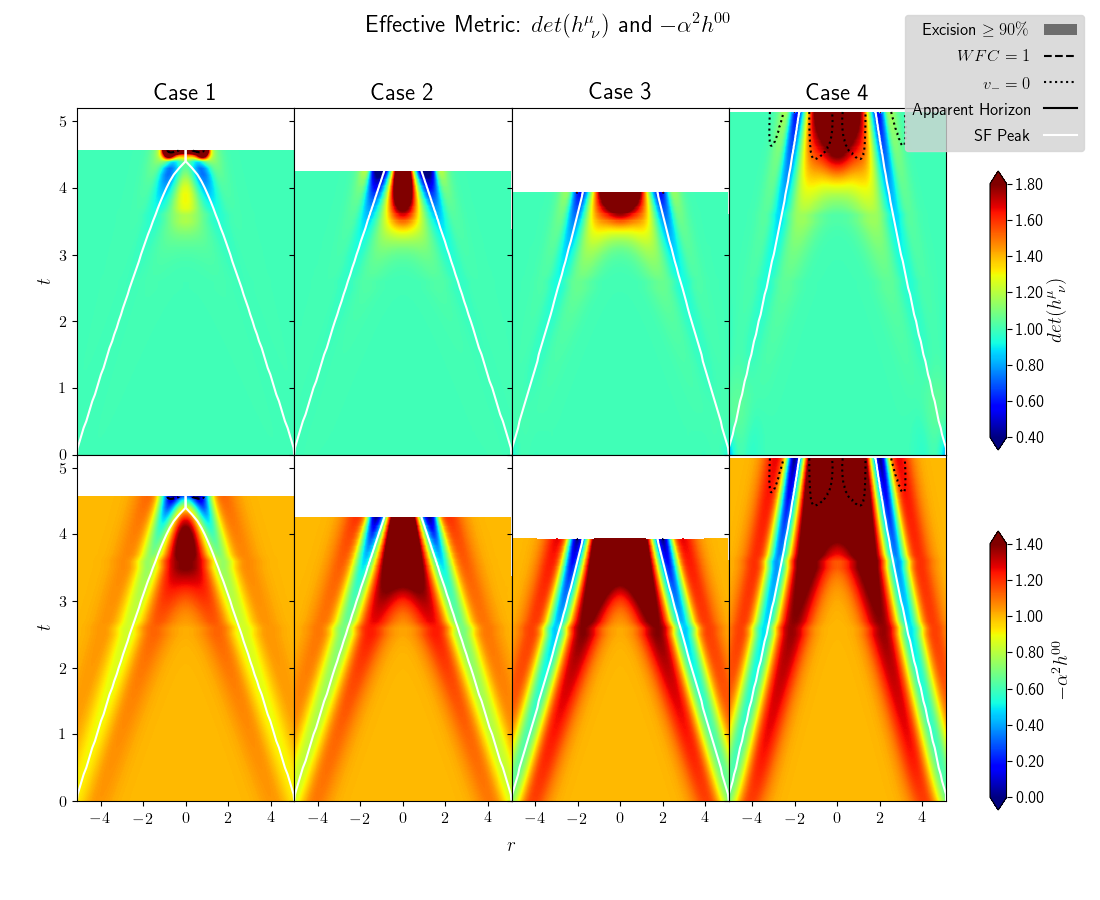}
\caption{$\det (h^\m_{~\n})$ (\textit{top}) and $-\a^2\, h^{00}$ (\textit{bottom}) for $G_3=g_3\,X$ with $g_3=0.4$. The corresponding values of the dimensionless coupling $\eta_3$, from left to right, are: $1.6\times 10^{-5},3.2\times 10^{-5},7\times 10^{-5}, 1\times 10^{-4}$. In all cases the evolution breaks down because $-\a^2h^{00}\to 0$, signalling a Keldysh-type-of transition.}
\label{fig:strong_g3_pos_det_h00}
\end{figure}

\begin{figure}[H]
\centering
\includegraphics[width=1\textwidth]{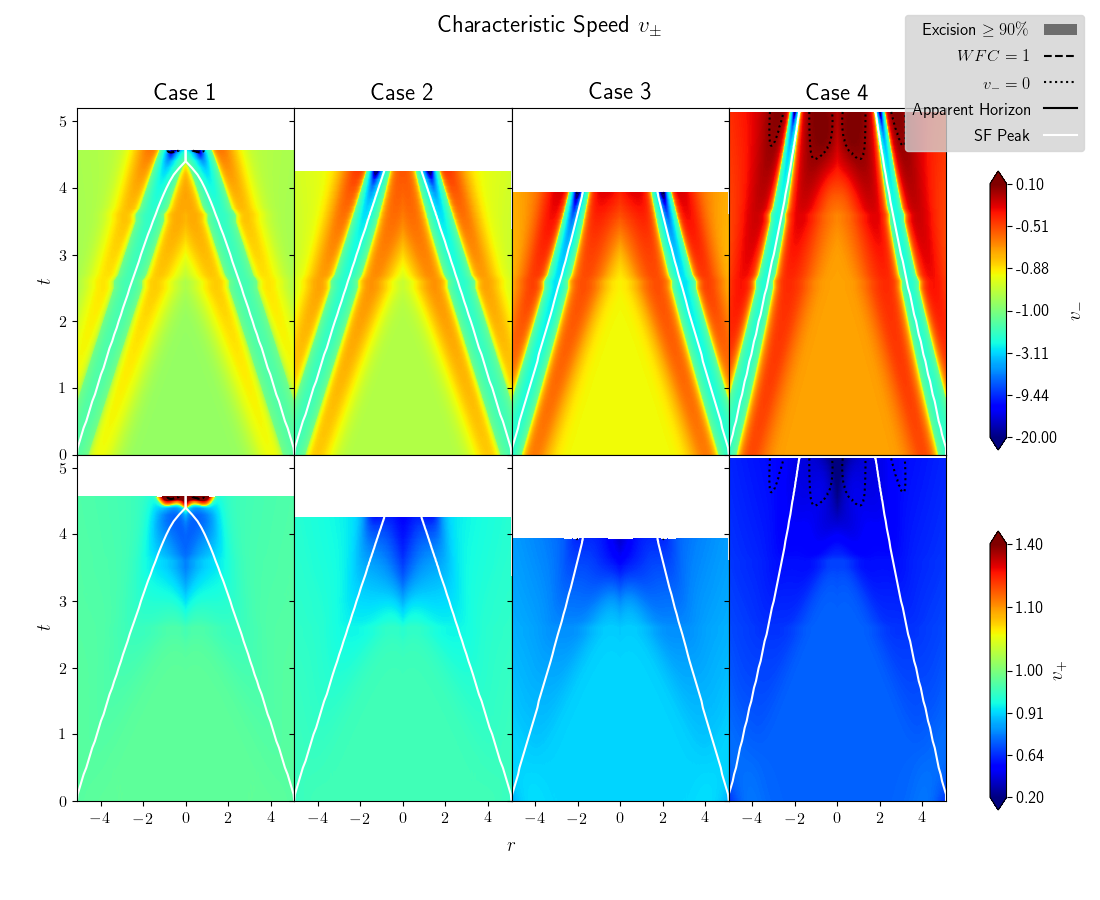}
\caption{Characteristic speeds of the outgoing (\textit{top}) and ingoing (\textit{bottom}) scalar modes for $G_3=g_3\,X$ with $g_3=0.4$. $v_-$ diverges at the transition, but $v_+$ remains finite.}
\label{fig:strong_g3_pos_speeds}
\end{figure}

In Figs. \ref{fig:intermdediate_g2_pos}, \ref{fig:strong_g3_pos_det_h00} and  \ref{fig:strong_g3_pos_speeds} of this Appendix we collect the results of some simulations that are relevant for the discussion in the main text. 

%

\clearpage
{
\bibliographystyle{JHEP}
\footnotesize
\bibliography{main}
}


\end{document}